\documentclass[12pt]{article}

\usepackage{scicite}

\usepackage{times}
\usepackage{graphicx}
\usepackage{amssymb}

\usepackage[hyphenbreaks]{breakurl}

\usepackage{hyperref}
\hypersetup{
    colorlinks=true,
    linkcolor=blue,
    filecolor=magenta,      
    urlcolor=cyan,
    pdftitle={Overleaf Example},
    pdfpagemode=FullScreen,
    }
\usepackage{xcolor}

\newenvironment{sciabstract}{%
\begin{quote} \bf}
{\end{quote}}
\newcommand\new[1]{{\color{black}#1}}          % new text for referee

\newcommand{\dotdeg}{\rlap{.}^\circ}
\newcommand{\dotarcsec}{\rlap{.}''}

\title{Giant Tidal Tails of Helium Escaping \\the Hot Jupiter HAT-P-32~b}

\author
{Zhoujian Zhang,$^{1,2,\ast}$ Caroline V. Morley,$^{2}$\\ Michael Gully-Santiago,$^{2}$ Morgan MacLeod,$^{3}$ Antonija Oklop{\v{c}}i{\'c},$^{4}$\\ Jessica Luna,$^{2}$ Quang H. Tran,$^{2}$ Joe P. Ninan,$^{5}$\\ Suvrath Mahadevan,$^{6,7,8}$ Daniel M. Krolikowski,$^{2,9}$ William D. Cochran,$^{10}$\\ Brendan P. Bowler,$^{2}$ Michael Endl,$^{11}$ Gudmundur Stef{\'a}nsson,$^{12}$\\ Benjamin M. Tofflemire,$^{2}$ Andrew Vanderburg,$^{13}$ Gregory R. Zeimann$^{14}$ \\
\\
\normalsize{$^{1}$Department of Astronomy \& Astrophysics, University of California, Santa Cruz, CA 95064, USA}\\
\normalsize{$^{2}$Department of Astronomy, The University of Texas at Austin, Austin, TX 78712, USA}\\
\normalsize{$^{3}$Center for Astrophysics, Harvard \& Smithsonian, Cambridge, MA 02138, USA}\\
\normalsize{$^{4}$Anton Pannekoek Institute for Astronomy, University of Amsterdam, The Netherlands}\\
\normalsize{$^{5}$Department of Astronomy \& Astrophysics, Tata Institute of Fundamental Research, India}\\
\normalsize{$^{6}$Department of Astronomy \& Astrophysics, The Pennsylvania State University, USA}\\
\normalsize{$^{7}$Center for Exoplanets and Habitable Worlds, USA}\\
\normalsize{$^{8}$ETH Zurich, Institute for Particle Physics \& Astrophysics, Zurich, Switzerland}\\
\normalsize{$^{9}$Steward Observatory, The University of Arizona, 933 N. Cherry Ave, Tucson, AZ 85721, USA}\\
\normalsize{$^{10}$Center for Planetary Systems Habitability and McDonald Observatory, UT Austin, USA}\\
\normalsize{$^{11}$McDonald Observatory and the Department of Astronomy, UT Austin, USA}\\
\normalsize{$^{12}$Princeton University, Department of Astrophysical Sciences, USA}\\
\normalsize{$^{13}$Department of Physics and Kavli Institute for Astrophysics and Space Research, MIT, USA}\\
\normalsize{$^{14}$Hobby-Eberly Telescope, The University of Texas at Austin, Austin, Austin, TX, 78712, USA}\\
\\
\normalsize{$^\ast$To whom correspondence should be addressed; E-mail:  zhangdirac [at] gmail [dot] com} 
}

\date{}

\begin{document} 


\baselineskip24pt


\maketitle


\begin{sciabstract}

\noindent Abstract:

Capturing planets in the act of losing their atmospheres provides rare opportunities to probe their evolution history. Such analysis has been enabled by observations of the helium triplet at 10833~\AA, but past studies have focused on the narrow time window right around the planet's optical transit. We monitored the hot Jupiter HAT-P-32~b using high-resolution spectroscopy from the Hobby-Eberly Telescope covering the planet's full orbit. We detected helium escaping HAT-P-32~b at a $14\sigma$ significance, with extended leading and trailing tails spanning a projected length over 53 times the planet's radius. These tails are among the largest known structures associated with an exoplanet. We interpret our observations using three-dimensional hydrodynamic simulations, which predict Roche Lobe overflow with extended tails along the planet's orbital path. 

\end{sciabstract}

\noindent {\bf Teaser: } \\
Long-baseline monitoring of the HAT-P-32Ab system reveals helium escaping through tidal tails 50 times the size of the planet.

% In setting up this template for *Science* papers, we've used both
% the \section* command and the \paragraph* command for topical
% divisions.  Which you use will of course depend on the type of paper
% you're writing.  Review Articles tend to have displayed headings, for
% which \section* is more appropriate; Research Articles, when they have
% formal topical divisions at all, tend to signal them with bold text
% that runs into the paragraph, for which \paragraph* is the right
% choice.  Either way, use the asterisk (*) modifier, as shown, to

%\paragraph*{Main Text} 

\section*{Introduction}

\noindent Atmospheric escape is the primary physical process sculpting the population of short-period, irradiated exoplanets. One piece of observational evidence of this process is the observed dearth of short-period Neptune-mass planets ({\it 1}). Several mechanisms likely contribute to the atmospheric escape, including photoevaporation and core-powered mass loss ({\it 2--9}), which predict distinct correlations between mass-loss rates and properties of the radiation environment (e.g., X-ray and ultraviolet fluxes of host stars) and planets (e.g., equilibrium temperatures). Directly measuring mass loss for a large ensemble of exoplanets can differentiate between these processes, and such measurements have been enabled by the helium 10833~\AA\ triplet ({\it 10--12}), which is a robust probe of exospheres accessible from the ground and immune to high absorption from the interstellar medium  which hampers similar studies based on Lyman~$\alpha$. 

Using the Habitable-zone Planet Finder Spectrograph (HPF; ({\it 13--15})) on the Hobby-Eberly Telescope (HET; ({16--18})), we observed HAT-P-32~b with time-series high-resolution ($R \approx 55,000$) spectra in order to detect a helium outflow and investigate atmospheric escape. HAT-P-32~b is a hot Jupiter transiting a late-F star, HAT-P-32~A, on a $2.15$-day orbit with an in-transit duration of $3.12$~hours ({\it 19--20}). This planet has an inflated radius ($1.79 \pm 0.03$~Jupiter radii) that almost fills its Roche lobe. A recent study ({\it 21}) discovered hydrogen and helium outflows escaping HAT-P-32~b using the Calar Alto high-Resolution search for M dwarfs with Exoearths with Near-infrared and optical \'{E}chelle Spectrographs (CARMENES; ({\it 22})) on the Calar Alto 3.5-m telescope. They collected high-resolution ($R \approx 80,000$) spectra of the host star over two nights right around the planet's optical transits and monitored for 6 hours each night centered on the middle of transits, leading to the detection of the planet's excess helium absorption with a maximum transit depth of $5.3\%$. 

With HET/HPF, we monitored HAT-P-32~Ab with time-series spectroscopy, covering orbital phases spanning the planet's full orbital period. We collected spectra during three planet transits on 2020 August 9 UT, September 19 UT, and October 6 UT, and during out-of-transit periods within two days of each transit. We also monitored the stellar activity of HAT-P-32~A with irregular cadence from 2020 August 1 UT to December 25 UT (Table~1). Our data were reduced using the HPF pipeline code \texttt{Goldilocks} and the \texttt{muler} python package ({\it 23--26}), which perform bias and non-linearity corrections, cosmic-ray rejection, flat fielding, wavelength calibration, and careful subtraction of sky emission features. We obtained a total of 77 high-quality spectra with a median signal-to-noise ratio (S/N) of $85$ per pixel ($142$ per resolution element) near 10833~\AA. All spectra were shifted to the stellar rest frame based on barycentric corrections and our computed absolute radial velocities (RVs).

\section*{Results}

\noindent We measured helium equivalent widths (EWs) for all HPF spectra of HAT-P-32~A+b and detected a long-duration (12~hours), significant (14$\sigma$) excess absorption feature near the transit of HAT-P-32~b (Fig. 1). The helium excess is not correlated with any stellar activity indicators (Fig.~S1) and spans $\approx 4\times$ longer than the planet's optical transit duration. To facilitate the transmission spectroscopic analysis, we divided the data into five subsets according to their orbital phases $\varphi$ (Fig.~1), which we denote as \textsc{start} ($-0.5 \leqslant \varphi \leqslant -0.15$; 23 spectra), \textsc{pre} ($-0.15 < \varphi \leqslant -0.03$; 6 spectra), \textsc{transit} ($-0.03 < \varphi < +0.03$; 18 spectra), \textsc{post} ($+0.03 \leqslant \varphi \leqslant +0.08$; 7 spectra), and \textsc{end} ($+0.08 < \varphi \leqslant +0.5$; 23 spectra). We constructed the out-of-transit reference spectra for each of the three transits by combining the corresponding \textsc{start} and/or \textsc{end} spectra (Fig. S2) and used them to normalize the other data to produce transmission and residual ($=$transmission$-1$) spectra. 

We detected strong excess helium absorption from \textsc{pre},  \textsc{transit}, and  \textsc{post} residual spectra and measured the time-dependent wavelength shift for each subset to characterize the velocity of HAT-P-32~b's escaping atmosphere (Fig.~2). We modeled the excess helium absorption feature in each residual spectrum (Figs.~S3--S4) using a Gaussian profile to determine its transit depth and central wavelength, and then we converted the latter into an RV of the escaping atmosphere (in the stellar rest frame) by comparing with the rest wavelength of the helium triplet. As shown in Fig. 2, RVs of the helium excess during the optical transits of HAT-P-32~b are consistent with those of the planet's orbital motion, reaffirming the planetary origin of the helium; the observed features have slightly higher RVs (i.e., toward the star) than the planet's orbital RV, which could result from mass transfer from the planet to its host star. These observations are consistent with both Roche lobe overflow and mass-loss controlled by planetary and stellar magnetic fields (e.g., {\it 27}). Before (\textsc{pre}) and after (\textsc{post}) the planet's optical transit, the observed escaping atmosphere does not track the planet's orbital motion and instead has only a small line-of-sight velocity shift in the stellar rest frame, suggesting the helium gas from the planet's upper atmosphere mostly moves within the sky plane perpendicular to the observers' line of sight; this property is consistent with the spatially extended geometry of the gas (as indicated by the long duration of our detected helium excess; see Fig.~1), with gas far from the planet orbiting the star. Helium excess features in \textsc{post} residual spectra are noticeably blue-shifted, implying that material trailing the planet is moving outward in the planetary system.

The maximum depth of the detected helium excess is measured to be $\approx 8.2\%$ during the optical transit of HAT-P-32~b; the helium excess in the 12-hour period surrounding optical transit is $5\%-6\%$ (Fig. 3). Our in-transit helium excess depth is about 1.5 times higher than the value ($5.3\%$) measured by CARMENES because their ``out-of-transit data'' were taken when the escaping helium is still in transit. Our detected excess helium absorption spans $3-4$~\AA\ in the stacked \textsc{transit} residual spectrum, comparable with the observed residual spectra from CARMENES ({\it 21}), and spans $1.5-2$~\AA\ in the stacked \textsc{pre} and \textsc{post} residual spectra. While other planets show trailing tails of material ({\it 12, 28--30}), HAT-P-32~b has both a (longer) leading and (shorter) trailing tail.

\section*{Discussion}

\noindent 

\noindent {\bf Helium outflow from three-dimensional hydrodynamic models}

\noindent Comparing our observations to three-dimensional hydrodynamic simulations of the HAT-P-32~A+b system provides physical insight into the geometry of the outflow. We generated three-dimensional hydrodynamic models following ({\it 31}) to examine the interactions between the planetary outflow and stellar winds in the tidal gravitational field of the HAT-P-32~A+b system (Fig.~4). Due to the small orbital separation and high star-to-planet mass ratio, HAT-P-32b nearly fills its Roche lobe. Our models show extended, columnar tails of planetary outflow both leading and trailing the planet along the orbital path. These tails provide excess helium absorption even at phases far from the planet's optical transit which match our observations. Our model also predicted that the mass loss rate of the planet is $\approx 1.7\times10^{-14}M_\odot$~yr$^{-1}$, implying the planet will lose its atmosphere over a timescale of $M_{\rm p}/\dot M_{\rm p}\approx 4\times10^{10}$~yr.

As shown in Fig.~1, our 3D model does not accurately explain the observed relative depth of the \textsc{pre} and \textsc{post} phases as compared to the mid-transit, which we used as a point of reference. However, as traced in Fig.~2, the model predicted that the extended tidal tails lie in approximately the stellar rest frame, not the planet's rest frame, as seen in our HPF data. Our simulations (Figs.~2 and 4) reveal how these extended tails span the star-planet environment of the HAT-P-32~A+b system. Sophisticated models that account for the momentum deposition by the stellar radiation field on the planetary outflow, orbital eccentricity of the planet, ram pressure of stellar wind, and the thermodynamics of the planetary outflow will be useful to probe the range of physical processes shaping this interaction.

\bigskip

\noindent {\bf The exceptional escaping helium of HAT-P-32~b}

\noindent HAT-P-32 b's escaping helium atmosphere is exceptional among known detections: it has the largest depth and longest duration found to date. The duration of the helium transit implies the tidal tails have a sky-projected length of 53 times the planet's radius (7 times the host star's radius), among the largest structures ever observed in a planetary system. Our study verifies the importance of long-baseline monitoring of planet-host systems to characterize systems with extended tails. Many surveys have targeted K-type planet-host stars since their UV spectra can readily populate the helium metastable state ({\it 32}); HAT-P-32 Ab's strong helium excess empirically demonstrates that F stars also provide a suitable environment for mechanisms triggering the planets' mass loss. Our observations of HAT-P-32b show that this planet-star configuration, where the planet largely fills its Roche lobe, can lead to extended outflowing material.

\bigskip

\noindent {\bf On the potential variability of HAT-P-32~b's helium excess}

\noindent We examined the potential variability of HAT-P-32~b's escaping helium atmosphere by comparing measured helium excess EWs across different dates (Fig.~5). During the planet's optical transit, the helium excess EWs mostly varied by less than $0.03$~\AA\ across three events. Near the planet's egress (observed during the second transit event) and slightly after the optical transit (shown by the POST subset), EWs of helium excess varied by less than $0.02$~\AA. In the PRE subset, all our observations were collected on the same night with the third transit event. In this subset, the measured helium excess EWs varied by $0.06$~\AA, and the helium excess absorption appears to become stronger with the increasing orbital phase. More observations of HAT-P-32~A+b over the orbital phase of the PRE subset will be useful to assess the variability of the planetary helium outflow slightly before the optical transit. To conclude, these HPF observations do not suggest a significant variability of HAT-P-32~b's helium excess over our monitoring baseline from 2020 August 7 to 2020 December 25. As a point of comparison, in our 3D hydrodynamic simulations, the overall column density of metastable helium and its mean radial velocity do not appear significantly variable despite instabilities at the interface between the planetary and stellar winds.

Comparing the helium excess revealed by the HPF spectra with those from the CARMENES data ({\it 21}), observed on 2018 September 1 and 2018 December 9, will examine the longer-term variability of the planet's helium excess. The published CARMENES spectroscopic observations span from about 4 hours before ($\approx -0.08$ in orbital phase) to 3 hours after ($\approx +0.06$ in orbital phase) the mid-point of the planet's optical transit, meaning they coincide with the PRE, TRANSIT, and POST subsets in our analysis (Fig.~1). New CARMENES observations that cover a longer time baseline (especially with orbital phases of our START and END subsets) than those already acquired by ({\it 21}) will construct the reference spectra that are needed to reanalyze the CARMENES-based helium excess. Comparing these results with the HPF measurements will investigate the long-term variability for HAT-P-32~b's outflow.

\bigskip

\noindent {\bf The diversity of planetary systems with escaping helium atmospheres} 

\noindent We compared properties of HAT-P-32~Ab with all other planetary systems that have either detections or upper-limit constraints of the escaping helium atmospheres, in order to investigate what physical parameters are causing the unusually extended helium atmosphere of HAT-P-32~b and driving the mass loss of exoplanets in general. We compiled all these systems in Table~S1. Following ({\it 12}), we computed the equivalent height of each planet's helium atmosphere $\delta{R_{p}}$ and its ratio to the planet's atmospheric scale height at the equilibrium temperature $H_{\rm eq}$, leading to a metric, $\delta R_{p}/H_{\rm eq}$, that quantifies the strength of the helium excess signal (also see {\it 33--35}). The census was divided into two subsets with planetary radii above and below $0.4$~$R_{\rm Jup}$, which represent gas giants and sub-Neptunes, respectively. In Fig.~6, we investigated $\delta R_{p}/H_{\rm eq}$ as functions of the planets' Roche-lobe filling, planetary surface gravity, the planet's bolometric equilibrium temperature, incident XUV flux from the host stars, and the host stars' effective temperatures. Here, the Roche-lobe filling stands for the ratio between the planetary radii $R_{p}$ to the Roche lobe radii $R_{\rm RL}$, with the latter computed via Equation~2 of ({\it 38}). 

With the existing census, Fig.~6 does not suggest clear trends between $\delta R_{p}/H_{\rm eq}$ and other physical properties investigated here. Among gas giants, HAT-P-32~b has the largest $\delta R_{p}/H_{\rm eq}$ and the longest-duration helium excess. This property lines up with its much larger Roche-lobe filling, lower surface gravity, and higher XUV incident flux, all of which are expected to foster the planet mass loss. However, several planets (e.g., WASP-69~b and WASP-107~b) with lower Roche-lobe fillings and XUV incident fluxes achieved comparable $\delta R_{p}/H_{\rm eq}$ as HAT-P-32~b. Also, WASP-76~b has very similar properties as HAT-P-32~b but its potential helium outflow was not significantly detected as shown by ({\it 39}), though partially due to the contamination of the telluric absorption near the helium triplet.

It is interesting to note that the three planets with the deepest helium excess depths, WASP-69~b, WASP-107~b, and HAT-P-32~b, also have longer-duration extended excess absorption than the planets' optical transits. A variety of system physical properties must conspire to create extended, deep helium signals; surface gravity---and therefore escape velocity---seems to be important since these three planets have some of the lowest in the sample. The observed diversity calls for further observational and theoretical studies of planets with both detections and non-detections of excess helium absorption with long-baseline monitoring, in order to probe the mass loss mechanisms of exoplanets.

\bigskip

\section*{Materials and Methods}

\noindent {\bf Known properties of the HAT-P-32 system}

\noindent Located at a distance of $283 \pm 2$~pc ({\it 40}), HAT-P-32~A is a moderately rapidly rotating ($v\sin{i} = 20.7 \pm 0.5$~km~s$^{-1}$) late-F dwarf with a near-solar metallicity ([Fe/H]$= -0.04 \pm 0.08$~dex) and an isochrone-based age of $2-4$~Gyr (e.g., {\it 19, 41}). The hot Jupiter HAT-P-32~b was discovered by ({\it 19}) from optical transit light curves and the host star's multi-epoch RVs. The observed RV jitter ($\approx 80$~m~s$^{-1}$) of HAT-P-32~A prevents precise measurements of the planet's orbital eccentricity, so ({\it 19}) provided two sets of orbital solutions for HAT-P-32~b by assuming a circular orbit (i.e., $e = 0$) or allowing the eccentricity to vary (leading to $e = 0.163 \pm 0.061$). Based on new RV data and/or transit light curves, ({\it 20}) and ({\it 42}) refined the planet's orbital eccentricity as $e=0.159^{+0.051}_{-0.028}$ and $e = 0.20^{+0.19}_{-0.13}$, respectively. However, the secondary eclipse timing of the planet detected by ({\it 43}) clearly suggests a lower $e = 0.007^{+0.070}_{-0.006}$. Throughout this work, we thus adopt a circular planet orbit for HAT-P-32~b. Various orbital analyses converge on the planet's orbital semi-major axis of $\approx 0.034$~au ({\it 19--20, 43}). HAT-P-32~b has a mass of $0.59 \pm 0.03$~$M_{\rm Jup}$, \new{a radius of $1.79 \pm 0.03$~$R_{\rm Jup}$, a low density of $0.14 \pm 0.02$~g~cm$^{-3}$, and a high equilibrium temperature of $1836 \pm 7$~K ({\it 19, 21}). With HAT-P-32~b's mass and irradiation, a core-less planet (the limiting case that provides the highest possible radius) would have a radius of $1.1-1.2$~$R_{\rm Jup}$ (e.g., ({\it \new{44}})), meaning that HAT-P-32~b is an inflated hot Jupiter.} This planet likely has a polar orbit, with a sky-projected obliquity of $85\dotdeg0 \pm 1\dotdeg5$, based on the Rossiter--McLaughlin effect ({\it \new{\new{45}}}). The HAT-P-32~A\new{+}b system also has an M1.5 stellar companion, HAT-P-32~B, with a projected separation of $2\dotarcsec94$ or $830$~au ({\it 43, \new{46--47}}). With such a wide orbit, HAT-P-32~B is not responsible for the detected significant RV trend ($-38 \pm 5$~m~s$^{-1}$~yr$^{-1}$) of HAT-P-32~A ({\it 20, \new{42}}).

\bigskip

\noindent \new{\bf HET/HPF observations, data reduction, and post processing}

\noindent HET is designed with a fixed elevation of $55^{\circ}$ and a tracking window of $\pm8^{\circ}$ ({\it 16--17}). As a consequence, HAT-P-32 (declination $=46\dotdeg687852$) is only observable over a track length of $1.5-3.3$~hours per night. The strategy of our survey is to observe the target over the full track when the planet's optical transit occurs and also collect out-of-transit spectra on both of the 2 nights before and 2 nights after the transit event to construct the reference spectra used to calibrate the transmission spectra. We also monitored HAT-P-32~A+b from 2020 August 1 UT to December 25 UT with irregular cadence to investigate its stellar activity. Our observing log is summarized in \new{Table~1}. 

Our raw HPF data were reduced using the \texttt{Goldilocks} pipeline ({\it 23--25}), which extracts the one-dimensional (1D) simultaneous spectra from the target fiber, sky reference fiber, and the laser frequency comb (LFC) fiber from two-dimensional images, spanning from 8100~\AA\ to 12800~\AA. Given the relaxed requirement about the radial velocity precision for the science goal of our survey and in order to avoid any scattered light in the target star fiber, the LFC was turned off. We applied standard 1D post-processing for these spectra using our newly developed \texttt{muler} framework ({\it 26}) which also provides quick-look quality assurance. 
 
To remove sky emission lines, \new{we combined the observed spectra from both the target and the sky fibers. These two types of fibers have different throughputs, so the direct/naive subtraction between target and sky fiber spectra is not sufficient. The sky-to-target throughput calibration was determined by our \texttt{muler} framework. Specifically, we analyzed twilight flats regularly acquired during the HPF operations. In this configuration, both target and sky fibers were illuminated by the scattered Sunlight. Computing the ratio of the solar spectra acquired by these two fibers, we thus derived a wavelength-dependent scaling factor that should be applied to the sky fiber before the sky subtraction. We found the sky fiber has a $\approx 7\%$ more throughput than the target fiber. We further validated this sky-to-target throughput calibration by using another dataset acquired during the HPF nighttime operations. In this dataset, both target and sky fibers were pointed at blank patches of the sky, leading to two sets of spectra populated with only sky emission lines. Comparing the strengths of these skylines from both fibers led to a similar calibration term determined from the twilight flats (also see the documentation and tutorials of \texttt{muler}). Overall, the sky subtraction based on \texttt{muler}} leads to a $14\times$ improvement over the naive sky subtraction, with negligible residual structure seen in most bright skylines near the helium triplet. \new{The performance of our sky subtraction is also verified by another independent method that can derive the sky-to-target throughput calibration by using a given target's observed spectrum without requiring twilight flats or blank sky observations ({\it \new{48}}).} \new{As an additional step to further minimize the impact of skylines on our analysis, we masked a skyline doublet which is close to the helium spectral feature; this doublet is located at 10832.103~\AA\ and 10832.412~\AA\ ({\it \new{49}}). For each spectrum, we first converted a wavelength range of 10832.103--10832.412~\AA\ into its stellar rest frame (see below) and then} used linear interpolation to approximate the fluxes. 
 
 Corrections to telluric absorption were not performed given we planned our observations strategically when telluric bands were widely separated from the helium spectral feature of HAT-P-32~Ab, as shifted by the Earth's barycentric motion (see Fig. 3). 
 
 Our reduced spectra are all in vacuum wavelengths. \new{All} spectra have signal-to-noise ratios (S/Ns) above 45 per pixel, i.e., $>75$ per resolution element, near the helium triplet at 10833~\AA; \new{the median S/N is 85 per pixel.} \new{During the observation, the stellar companion B is outside the HPF target fiber, the radius of which ($0.85''$) is only $29\%$ of the angular separation between A and B components ($2.94''$; {\it \new{46--47}}). Also, the B-to-A flux ratio is about 0.017 near the helium triplet at 10833~\AA\ (see Figure~4 of {\it \new{43}}); this ratio is comparable to the noise-to-signal ratio of A's observed fluxes, with a median of 0.012 ($=1/85$). Thus, the stellar companion would have comparable fluxes with A's flux uncertainties, even if both stellar components were observed by the same fiber. Therefore, the stellar companion HAT-P-32~B has negligible contamination to our observations of HAT-P-32~A+b. }

\bigskip

\noindent \new{\bf HPF relative radial velocities of HAT-P-32~A}

\noindent We measured the relative RVs of HAT-P-32~A from the HET/HPF spectra following ({\it \new{50}}). We first applied the barycentric correction to all spectra using Astropy ({\it \new{51--52}}) and then multiplied each spectrum by a third-order polynomial, with coefficients determined in the steps described below, to match the fluxes of the science spectrum (observed on 2020 September 19 UT) with the highest S/N among our entire dataset. To create a master template, we carried out a cubic basic (B-spline) regression to data points from all these scaled spectra via the least-squares minimization implemented with a $\kappa$-sigma clipping that removes significant outliers from the residuals. We then jointly fitted the relative RV and the third-order polynomial coefficients for each spectral order of each spectrum by minimizing the $\chi^{2}$ between the given spectrum and the master template, over a grid of velocities (from $-5$~km~s$^{-1}$ to $+5$~km~s$^{-1}$ with steps of $50$~m~s$^{-1}$) following ({\it \new{53}}). During this process, we also masked the telluric absorption and sky emission features. We determined the RV value and variance of a given spectral order based on the $\chi^{2}$--velocity parabola and computed the final relative RV for each spectrum as the weighted mean and standard error of RVs among all spectral orders. The spectrum with the highest S/N in our dataset thus provides the baseline for the measured relative RVs. Our resulting HPF relative RVs have typical uncertainties of $131$~m~s$^{-1}$. 

Given that HAT-P-32~A has an absolute RV of $-23.21 \pm 0.26$~km~s$^{-1}$ based on ({\it 19}), we thus added this value to the relative RV of each spectrum to obtain an absolute RV, with uncertainties added in quadrature. The computed absolute RVs of our data have a typical uncertainty below $0.3$~km~s$^{-1}$, corresponding to a wavelength shift of only $<0.01$~\AA\ near the helium triplet. We used the computed absolute RV of each spectrum to shift it into the stellar rest frame.

\bigskip

\noindent \new{\bf Equivalent width measurements of helium and the calcium infrared triplets}

\noindent We measured the equivalent widths (EWs) of the helium triplet in the stellar rest frame by integrating the line flux over $10831.5-10834.5$~\AA, with the pseudo-continuum approximated by a linear fit of the fluxes from two surrounding wavelength regions of $10824-10826$~\AA\ and \new{$10840-10841$~\AA}. The flux uncertainties are propagated into our resulting EWs in an analytical fashion. We also measured the EWs of the Calcium infrared triplet (Ca~\textsc{ii} IRT) at $8500$~\AA, $8544$~\AA, and $8644$~\AA, which probe the stellar chromospheric activity. These lines are thus used to validate that our observed significant excess helium absorption has a planetary rather than stellar origin (e.g., ({\it 21, \new{54}})). To compute the EW of each triplet component, we defined the line wavelength as $8498 - 8503$~\AA\ with the pseudo-continuum established from $8491-8495$~\AA\ and $8505-8509$~\AA\ for Ca~\textsc{ii}~$\lambda\lambda 8500$; line wavelength as $8540-8549$~\AA\ with the pseudo-continuum from $8536-8537.5$~\AA, $8552-8556$~\AA, and $8562-8570$~\AA\ for Ca~\textsc{ii}~$\lambda\lambda 8545$; and line wavelength as $8661-8668$~\AA\ with the pseudo-continuum from $8658-8660$~\AA, $8672-8676$~\AA\ and $8680-8688$~\AA\ for Ca~\textsc{ii}~$\lambda\lambda 8665$. For a given spectrum, we adopted the EW of the Ca~\textsc{ii}~IRT as the sum of all three components' EWs. As shown in Fig. S1, our measured Ca~IRT EWs of spectra in \textsc{pre}, \textsc{transit}, and \textsc{post} subsets, when the escaping helium atmosphere and/or the planet are in transit, are all consistent with those of out-of-transit data in \textsc{start} and \textsc{end} subsets. Therefore, the observed helium excess spectral features likely originate from the planet's upper atmosphere. This conclusion is supported by the observed low activity level from the ground-based optical light curves (e.g., ({\it \new{55}})) and was also drawn by ({\it 21}) based on CARMENES data.

\bigskip

\noindent \new{\bf Construction of the reference, transmission, and residual spectra}

\noindent We re-sampled each spectrum in the stellar rest frame to the same wavelength grid of the spectrum that has the highest S/N at 10833~\AA. Then we normalized the spectrum in the neighborhood of the helium triplet, with the pseudo-continuum established from a linear fit of fluxes from $10824-10826$~\AA\ and \new{$10840-10841$}~\AA\ as used in our EW measurements. For data collected from each of the three transit events and also from the long-term stellar activity monitoring, we derived a reference spectrum by computing the average of all spectra corresponding to orbital phases below $-0.15$ (i.e., \textsc{start}) or above $+0.08$ (i.e., \textsc{end}). We propagated the flux uncertainties in an analytical fashion and obtained a total of four reference spectra. As shown in Fig. S2, our computed reference spectra all have consistent shapes and values in the neighboring wavelengths of the helium triplet, with different fluxes from the telluric absorption lines that are clearly separated from the helium feature. We then divided each normalized \textsc{pre}, \textsc{transit}, and \textsc{post} spectrum by its corresponding reference spectrum to produce transmission and residual ($=$ transmission$-1$) spectra (Figs. S3--S4). \new{A telluric band, spanning 10835--10840~\AA\ in the stellar rest frame, is near but separated from the helium feature in our residual spectra. Thus, our analysis cannot detect if there is any helium excess from a clump of the planet's escaping atmosphere with an RV toward the host star with values of $48-187$~km~s$^{-1}$; such signal is not predicted by our subsequent three-dimensional hydrodynamic simulation for our target (see below). }

\bigskip

\noindent \new{\bf Gaussian models of the excess helium absorption}

\noindent The observed wavelengths of helium excess reflect the RVs of helium upper atmosphere escaping HAT-P-32~b. In order to identify these wavelengths, we fitted a Gaussian profile $G(\lambda) = -A_{G}\ {\rm \exp{\left[- (\lambda - \mu_{G})^{2}/2\sigma_{G}^{2}\right]}}$ to each residual spectrum over $10825 - 10840$~\AA, with $A_{G}$, $\mu_{G}$, and $\sigma_{G}$ describing the transit depth, central wavelength, and standard deviation wavelength of each excess absorption feature. We excluded telluric bands red-ward of the helium triplet, as well as wavelengths of sky emission lines whose fluxes were replaced by linear interpolation during the data processing. Some residual spectra observed near the end of the planet's optical transit show broad absorption features that have slightly shorter wavelengths and shallower depths than the helium excess (e.g., the residual spectrum of the second transit event with the orbital phase near 0.012; Fig. S3). These features line up with the weakest component of the helium triplet (Fig. 3) but contain an absorption component that is even more blue-shifted (also see ({\it 21})). We chose to exclude wavelengths of such feature in our fitting process given that the goal of fitting Gaussian profiles is to identify wavelengths of the primary component of the excess helium absorption. 

Based on our fitted $A_{G}$ values, the depth of helium excess from each residual spectrum in the \textsc{pre} (6 spectra in total), \textsc{transit} (18 spectra in total), and \textsc{post} (7 spectra in total) subsets spans $2\%-3\%$, $4\%-7\%$, and $2\%-4\%$, respectively. The transit depth in each subset is further increased after stacking all residual spectra (Fig. 3). Also, our observed helium excess has full widths at half maximum of $1.5-3$~\AA\ (i.e., $2.355\sigma_{G}$), comparable with the detected helium excess in other planetary systems (e.g., ({\it 11--12})). To compute the RV of the escaping helium atmosphere of HAT-P-32~b, we further compared the fitted central wavelength $\mu_{G}$ of each residual spectrum with the mean rest wavelength of the strongest two components of the helium triplet at 10833.26~\AA. Our fitted $\mu_{G}$, $\sigma_{G}$, and computed escaping RVs are presented in Fig. 2.

\bigskip

\noindent \new{\bf Planetary orbital RV in the stellar rest frame}

\noindent We computed the planet's RV in the stellar rest frame as a function of the planet's orbital phase ($\varphi$) based on the following equation,
\begin{displaymath}
RV_{p} (\varphi) = - \left(1 + \frac{1}{M_{p}/M_{\star}}\right) \times K_{\star} \left[ \cos{\left(\nu(\varphi) + \omega_{\star}\right)} + e\cos{\omega_{\star}} \right] \quad \quad \quad {\rm (Eq.\ S1)}
\end{displaymath}
where $M_{p}/M_{\star}$ is the planet-to-star mass ratio, $K_{\star}$ is the semi-amplitude of the host star's RV, $\omega_{\star}$ is the argument of the periastron of the host star's orbit induced by the exoplanet, and $e$ is the planet's orbital eccentricity. The true anomaly, $\nu$, is converted from a given orbital phase based on the planet's orbital period ($P$) and eccentricity ($e$), time of periastron ($T_{P}$), and the time of the planet's optical transit ($T_{C}$).

We computed $RV_{p}(\varphi)$ for a circular planet orbit and adopted $M_{p}/M_{\star} = 4.81 \times 10^{-4}$ ({\it 21}), $K_{\star} = 83.4$~m~s$^{-1}$ ({\it 21}), $P = 2.1500082$~days ({\it 20}), and $T_{C} = 2456237.031$~BJD$_{\rm TDB}$ ({\it 20}). We simply assumed $\omega_{\star} = 0^{\circ}$ and $T_{P} = T_{C} - P/4$ for using Eq. S1. We did not incorporate the uncertainties of orbital parameters in our calculation, given our analysis compares the planet's RV with measured RVs of the escaping helium atmosphere under a qualitative perspective. Our computed $RV_{p}(\varphi)$ is presented in Fig. 2.

\bigskip

\noindent \new{\bf Stacked residual spectra and the light curve of excess helium absorption}

\noindent We shifted all residual spectra to an ``escaping atmospheric rest frame'' based on the computed RVs of the helium upper atmosphere escaping HAT-P-32~b, \new{such that the helium excess feature in each shifted residual spectrum all lines up with the rest wavelength of the strongest component of the helium triplet at 10833.26~\AA}. We then computed the stacked residual spectrum based on the weighted flux mean and uncertainty of all data in each of the \textsc{pre}, \textsc{transit}, and \textsc{post} subsets (Fig. 3). In addition to the excess helium absorption features, the stacked \textsc{pre} residual spectrum shows a deep absorption feature near 10835~\AA, which is caused by the telluric absorption mainly contributed by the \textsc{pre} residual spectrum with an orbital phase of $-0.098$ and $-0.111$ (Fig. S4). The wavelength of this feature is outside the wavelength range used to compute the helium equivalent widths (Fig. 1). To compute the light curve of excess helium absorption (Fig. 1), we integrated each residual spectrum in the escaping atmospheric rest frame over a window centered at $10833.26$~\AA\ (the mean rest wavelength of the strongest two components of the helium triplet) with a width of $\pm 1.8$~\AA\ (corresponding to an RV range of $\pm 50$~km~s$^{-1}$), with spectral uncertainties propagated in an analytical fashion.

\bigskip

\noindent \new{\bf Three-dimensional hydrodynamic models} 

\noindent Our 3D dydrodynamic models made use of the Athena++ (magneto)hydrodynamics code ({\it \new{56}}), which employs an Eulerian algorithm with active and static mesh refinement capabilities. Our models simulated the properties of the interaction between the planetary outflow (with properties based on parameterized assumptions) and the stellar wind and the planet-star gravitational field (e.g., {\it 31}). We adopted a nearly-isothermal ideal-gas equation of state with an adiabatic index $\gamma=1.0001$ (allowing each of the stellar and planetary outflows to be nearly isothermal but with very different temperatures) and ignored any possible effects of stellar or planetary magnetic and radiation fields. 

We solved the hydrodynamics equations in a rotating reference frame centered on the host star HAT-P-32~A. Our frame rotates with the planetary mean motion to minimize the orbital advection of planetary outflow across the coordinate mesh. We employed a spherical-polar mesh that covers the full $4\pi$ of solid angle and extends in radius from HAT-P-32~A's stellar radius of $R_{\star} \approx 9.5 \times 10^{10}$~cm (\new{or $1.37 \pm 0.03$~$R_{\odot}$ as determined by} {\it 20}), to about $100 \times R_{\star}$ of $9\times10^{12}$~cm. \new{With a semi-major axis of 0.034~au~$\sim 5\times10^{11}$~cm ({\it 19--20, \new{43}}) or 5.3~$R_{\star}$, the orbit of HAT-P-32~b is covered by this coordinate mesh.} Our base mesh employs $144\times96\times192$ zones in the $(r,\theta,\phi)$ coordinate, with the planet placed at $\theta = \pi/2$. To increase the resolution of simulations \new{near the planet's orbit}, we add static mesh refinement. Specifically, we refined an equatorial torus extending $2\times10^{11} - 8\times10^{11}$~cm in $r$, $\pi/4 - 3\pi/4$ in $\theta$, and $0 - 2\pi$ in $\phi$ by one level above the base mesh, \new{or a factor of two higher spatial resolution}. We also refined a box surrounding the planet location on the mesh by four levels of refinement \new{(i.e., $16\times$ higher spatial resolution than the base mesh)} to capture the planet-scale outflow with sufficient resolution in our simulation. 

We set our model parameters based on orbital properties (with $e = 0$) that have been derived for HAT-P-32~b ({\it 19, 21}). To model the planetary outflow, we used the hydrodynamic escape parameter, 
\begin{displaymath}
    \lambda = \frac{G M_{\rm p}}{R_{\rm p} c_s^2} \quad \quad \quad {\rm (Eq.\ S2)}
\end{displaymath}
where $M_{\rm p}$ is the planet's mass, $R_{\rm p}$ is the planet radius, and $c_s$ is the sound speed of the outflow. We adopted $\lambda=8$ for the planetary wind in our models, implying an outflow temperature of $\approx 5750$~K. This is a free parameter of our current models and we find that $\lambda$ particularly affects the stream kinematics, as well as the inferred planetary mass loss rate ({\it \new{57}}). To model the much hotter ($\approx 10^6$~K), fast-expanding stellar wind, we adopted $\lambda = 15$ and replace $M_{\rm p}$ and $R_{\rm p}$ to be the mass and radius of the host star, respectively, in the Equation S2. \new{The planetary mass loss rate in our model is $1.07\times10^{12}$~g~s$^{-1}$ ($\approx 1.7\times10^{-14}M_\odot$~yr$^{-1}$). This implies a loss timescale for the planetary envelope of $M_{\rm p}/\dot M_{\rm p} \approx3.8\times 10^{10}$~yr.} The stellar mass loss rate in our model, $7\times10^{-14}M_\odot$~yr$^{-1}$, is on the order of typical main sequence mass loss rates ({\it \new{58}}).

We post-processed these snapshots by iterating to find the stellar and planetary outflow ionization states in the stellar radiation field. \new{To construct the spectral energy distribution of the host star, we combined the published X-ray and ultraviolet (XUV) spectrum of $\tau$~Boo ($0-1200$~\AA; {\it 59}), which has the same spectral type as HAT-P-32~A, and the BT-Settl model spectrum ($>1200$~\AA) with an effective temperature of 6300~K and logarithmic surface gravity of $\log{(g)} = 4.5$~dex; we further scaled this spectrum based on the XUV flux estimates from ({\it 21}).} We then cast rays from the star through the domain to an observer. The full methodology is presented in ({\it 31}).

Fig.~4 demonstrates a slice of gas density through the simulation model domain centered on HAT-P-32~A, showing that an outflow from HAT-P-32~b extends nearly along the orbital path both leading and trailing the planet. This outflow is shaped by its interactions with the much-faster stellar wind (as is visible from unstable interfaces between the higher-density planetary outflow and the lower-density stellar wind). However, the primary force shaping the planetary outflow is the tidal gravity of the star-planet system, which is particularly strong for HAT-P-32~b because the planet is nearly filling its Roche lobe ({\it 21}). As a result, the planetary outflow is slow-moving by comparison to the planet's orbital speed. The differential orbital frequency as a function of the distance from the star means that gas in the planetary outflow is advected into leading and trailing arms of the planet's upper atmosphere. 

The consequences for the observable properties of excess helium absorption of this outflow geometry are dramatic. \new{Fig.~7} presents the number density of metastable helium and the radial velocity of gas in the region surrounding the planet. Overlying contours show the cumulative, radial optical depth from the host star in the metastable helium line. These contours show that the columnar structure of planetary outflow maintains relatively high optical depth even at large distances from the planet. This can be contrasted to a spherical outflow, in which the optical depth drops steeply with distance above the planetary surface. For the observable properties of HAT-P-32~b's transit, this means that the extended absorption in \textsc{pre} and \textsc{post} phases (e.g., Fig.~1) can be attributed to columns of planetary outflow that are strongly shaped by orbital advection into leading and trailing tails. 

\new{Fig.~7} also shows the radial velocity of material with the same overlying contours of optical depth. We see that on the leading arm, planetary material has a slight redshift, implying flow toward the host star, while material in the trailing arm is slightly blue-shifted. Most importantly, the radial velocities of gas are low compared to the planet's orbital velocity ($\sim 170$~km~s$^{-1}$). The leading and trailing tails of the planet's escaping atmosphere, therefore, lie close to the stellar rest frame as most of its motion is along the orbit, perpendicular to an observer during transit. Several properties of the model affect the kinematics of the tidal tails of planetary mass loss. We find that one of the strongest effects is the characteristic temperature of the planetary outflow (controlled by the parameter $\lambda$ in our models). Cooler tidal tails are shaped more strongly by the tidal potential and are broadened less by their own thermal expansion. This suggests that more sophisticated modeling that fully reproduces the equivalent width light curve (Fig. 1) and kinematics as a function of phase (Fig. 2) will be very constraining about the planetary outflow's thermodynamics. 

The \textsc{pre} and \textsc{post} tails of material need not share an identical temperature (as they currently do in our hydrodynamic models). We find that cooler \textsc{post} tails better reproduce the redshifts observed several hours after mid-transit. Additionally, it is important to highlight that there are physical processes not included in our models that could influence tidal tail kinematics. We have not included momentum deposition by the stellar radiation field on the outflow directly in our models, nor have we systematically varied the orbital eccentricity, the relative strength of the stellar wind, or the thermodynamics of the planetary outflow. \new{Stellar wind and radiation pressure act similarly, applying a roughly $r^{-2}$-scaled radial pressure away from the star. Increasing these effects blueshifts the trailing tail, and can erode the leading tail until the outflow forms a cometary morphology ({\it 31}). Eccentricity introduces waves into the tails, perhaps including subtle radial velocity signatures. Because these effects interact nonlinearly, sophisticated hydrodynamic simulations are needed to explore this phase space and to determine to what degree the observational constraints uniquely determine HAT-P-32~A+b's properties. 

We also supplemented our 3D modeling with 1D hydrodynamic models (following ({\it \new{60--61}}) of planetary outflows of varying temperatures, with otherwise equivalent assumptions. At identical temperatures and equivalent width, these 1D models predict $\sim 2 \times$ higher planetary mass loss rate because they neglect the compression of material into tidal tails. The 1D models also suggest that the best-fitting mass-loss rate scales as $\dot M_p \propto T$ (i.e., $\dot M_p \propto \lambda_p^{-1}$, where $T$ is the outflow temperature), thus giving an indication of how varying the uncertain planetary outflow thermodynamics affects the inferred planetary mass loss rate (also see ({\it \new{57}})). }

% Your references go at the end of the main text, and before the
% figures.  For this document we've used BibTeX, the .bib file
% scibib.bib, and the .bst file Science.bst.  The package scicite.sty
% was included to format the reference numbers according to *Science*
% style.

%BibTeX users: After compilation, comment out the following two lines and paste in
% the generated .bbl file. 

\newpage

%\bibliography{scibib}
%\bibliographystyle{Science}

%%%%%%%%%%
% Bibliography
%%%%%%%%%%

\section*{References and Notes} 

\begin{enumerate} 

\item Mazeh, T., \new{Holczer, T., \& Faigler, S.,} Dearth of short-period Neptunian exoplanets: A desert in period-mass and period-radius planes, {\it A\&A}, {\bf 589}, 75\new{--81} (2016)

\item Owen, J. E., \& Wu, Y., Kepler Planets: A Tale of Evaporation, {\it ApJ}, {\bf 775}, 105\new{--116} (2013)

\item  Lopez, E. D., \& Fortney, J. J., The Role of Core Mass in Controlling Evaporation: The Kepler Radius Distribution and the Kepler-36 Density Dichotomy, {\it ApJ}, {\bf 776}, 2\new{--12} (2013)

\item  Owen, J. E., \& Wu, Y., The Evaporation Valley in the Kepler Planets, {\it ApJ}, {\bf 847}, 29\new{--42} (2017)

\item  Ginzburg, S., \new{Schlichting, H.~E., \& Sari, R.}, Core-powered mass-loss and the radius distribution of small exoplanets, {\it MNRAS}, {\bf 476}, 759\new{--765} (2018)

\item  Gupta, A., \& Schlichting, H. E., Sculpting the valley in the radius distribution of small exoplanets as a by-product of planet formation: the core-powered mass-loss mechanism, {\it MNRAS}, {\bf 487}, 24\new{--33} (2019)

\item Gupta, A., \& Schlichting, H. E., Signatures of the core-powered mass-loss mechanism in the exoplanet population: dependence on stellar properties and observational predictions, {\it MNARS}, {\bf 493}, 792\new{--806} (2020)

\item Loyd, R.~O.~P., \new{{Shkolnik}, E. L., {Schneider}, A. C., {Richey-Yowell}, T., {Barman}, T. S., {Peacock}, S., {Pagano}, I.,} Current Population Statistics Do Not Favor Photoevaporation over Core-powered Mass Loss as the Dominant Cause of the Exoplanet Radius Gap, {\it ApJ}, {\bf 890}, 23\new{--43} (2020)

\item Rogers, J. G., \new{{Gupta}, A., {Owen}, J. E., {Schlichting}, H. E.}, Photoevaporation versus core-powered mass-loss: model comparison with the 3D radius gap, {\it MNRAS}, {\bf 508}, 5886\new{--5902} (2021)

\item Spake, J. J., \new{{Sing}, D.~K., {Evans}, T.~M., {Oklop{\v{c}}i{\'c}}, A., {Bourrier}, V., {Kreidberg}, L., {Rackham}, B.~V., {Irwin}, J., {Ehrenreich}, D., {Wyttenbach}, A., {Wakeford}, H.~R., {Zhou}, Y., {Chubb}, K.~L., {Nikolov}, N., {Goyal}, J.~M., {Henry}, G.~W., {Williamson}, M.~H., {Blumenthal}, S., {Anderson}, D.~R., {Hellier}, C., {Charbonneau}, D., {Udry}, S., {Madhusudhan}, N.}, Helium in the eroding atmosphere of an exoplanet, {\it Nature}, {\bf 557}, 68\new{--70} (2018)

\item Allart, R., \new{{Bourrier}, V., {Lovis}, C., {Ehrenreich}, D., {Spake}, J.~J., {Wyttenbach}, A., {Pino}, L., {Pepe}, F., {Sing}, D.~K., {Lecavelier des Etangs}, A.}, Spectrally resolved helium absorption from the extended atmosphere of a warm Neptune-mass exoplanet, {\it Science}, {\bf 362}, 1384\new{--1387} (2018)

\item Nortmann, L., \new{{Pall{\'e}}, E., {Salz}, M., {Sanz-Forcada}, J., {Nagel}, E., {Alonso-Floriano}, F. J., {Czesla}, S., {Yan}, F., {Chen}, G., {Snellen}, I.~A.~G., {Zechmeister}, M., {Schmitt}, J.~H.~M.~M., {L{\'o}pez-Puertas}, M., {Casasayas-Barris}, N., {Bauer}, F. F., {Amado}, P. J., {Caballero}, J. A., {Dreizler}, S., {Henning}, T., {Lamp{\'o}n}, M., {Montes}, D., {Molaverdikhani}, K., {Quirrenbach}, A., {Reiners}, A., {Ribas}, I., {S{\'a}nchez-L{\'o}pez}, A., {Schneider}, P. C., {Zapatero Osorio}, M. R.}, Ground-based detection of an extended helium atmosphere in the Saturn-mass exoplanet WASP-69b, {\it Science}, {\bf 362}, 1388\new{--1391} (2018)

\item Mahadevan, S., \new{{Ramsey}, L., {Bender}, C., {Terrien}, R., {Wright}, J. T., {Halverson}, S., {Hearty}, F., {Nelson}, M., {Burton}, A., {Redman}, S., {Osterman}, S., {Diddams}, S., {Kasting}, J., {Endl}, M., {Deshpande}, R.}, The habitable-zone planet finder: a stabilized fiber-fed NIR spectrograph for the Hobby-Eberly Telescope, {\it SPIE}, vol. 8446, p. 84461S, \new{14 pages} (2012)

\item Mahadevan, S., \new{{Ramsey}, L. W., {Terrien}, R., {Halverson}, S., {Roy}, A., {Hearty}, F., {Levi}, E., {Stefansson}, G. K., {Robertson}, P., {Bender}, C., {Schwab}, C., {Nelson}, M.}, The Habitable-zone Planet Finder: A status update on the development of a stabilized fiber-fed near-infrared spectrograph for the for the Hobby-Eberly telescope, {\it SPIE}, vol. 9147, p. 91471G, \new{10 pages} (2014)

\item Mahadevan, S., \new{Anderson, T., Balderrama, E., Bender, C. F., Bevins, E., Blakeslee, S., Cole, A., Conran, D., Diddams, S., Dykhouse, A., Darling, J., Fredrick, C., Halverson, S., Hearty, F., Jennings, J. Kaplan, K., Kanodia, S., Levi, E., Lubar, E., Metcalf, A. J., Monson, A., Ninan, J., Nitroy, C., Ramsey, L., Robertson, P., Roy, A., Schwab, C., Shetrone, M., Spencer, R., Stefansson, G., Terrien, R., Wright, J.}, The habitable-zone planet finder: engineering and commissioning on the Hobby Eberly telescope (Conference Presentation), {\it Ground-based and Airborne Instrumentation for Astronomy VII}, vol. 10702, p. 1070214 (2018)

\item Ramsey, L. W., \new{{Ramsey}, L. W., {Adams}, M.~T., {Barnes}, T. G., {Booth}, J. A., {Cornell}, M. E., {Fowler}, J. R., {Gaffney}, N. I., {Glaspey}, J. W., {Good}, J. M., {Hill}, G. J., {Kelton}, P. W., {Krabbendam}, V. L., {Long}, L., {MacQueen}, P. J., {Ray}, F. B., {Ricklefs}, R. L., {Sage}, J., {Sebring}, T. A., {Spiesman}, W.~J., {Steiner}, M.}, Early performance and present status of the Hobby-Eberly Telescope, {\it SPIE}, vol. 3352, p. 34\new{--42} (1998)

\item Shetrone, M., \new{{Cornell}, M. E., {Fowler}, J. R., {Gaffney}, N., {Laws}, B., {Mader}, J., {Mason}, C., {Odewahn}, S., {Roman}, B., {Rostopchin}, S., {Schneider}, D. P., {Umbarger}, J., {Westfall}, A.}, Ten Year Review of Queue Scheduling of the Hobby-Eberly Telescope, {\it PASP}, {\bf 119}, 855, \new{556--566} (2007)

\item Hill, G. J., \new{{Lee}, H., {MacQueen}, P. J., {Kelz}, A., {Drory}, N., {Vattiat}, B. L., {Good}, J. M., {Ramsey}, J., {Kriel}, H., {Peterson}, T., {DePoy}, D.~L., {Gebhardt}, K., {Marshall}, J.~L., {Tuttle}, S. E., {Bauer}, S. M., {Chonis}, T. S., {Fabricius}, M. H., {Froning}, C., {H{\"a}user}, M., {Indahl}, B. L., {Jahn}, T., {Landriau}, M., {Leck}, R., {Montesano}, F., {Prochaska}, T., {Snigula}, J. M., {Zeimann}, G., {Bryant}, R., {Damm}, G., {Fowler}, J.~R., {Janowiecki}, S., {Martin}, J., {Mrozinski}, E., {Odewahn}, S., {Rostopchin}, S., {Shetrone}, M., {Spencer}, R., {Mentuch Cooper}, E., {Armandroff}, T., {Bender}, R., {Dalton}, G., {Hopp}, U., {Komatsu}, E., {Nicklas}, H., {Ramsey}, L. W., {Roth}, M. M., {Schneider}, D. P., {Sneden}, C., {Steinmetz}, M.}, The HETDEX Instrumentation: Hobby-Eberly Telescope Wide-field Upgrade and VIRUS, {\it AJ}, {\bf 162}, 298\new{--332} (2021)

\item Hartman, J. D., \new{{Bakos}, G. {\'A}., {Torres}, G., {Latham}, D.~W., {Kov{\'a}cs}, G., {B{\'e}ky}, B., {Quinn}, S.~N., {Mazeh}, T., {Shporer}, A., {Marcy}, G.~W., {Howard}, A.~W., {Fischer}, D.~A., {Johnson}, J.~A., {Esquerdo}, G.~A., {Noyes}, R.~W., {Sasselov}, D.~D., {Stefanik}, R.~P., {Fernandez}, J.~M., {Szklen{\'a}r}, T., {L{\'a}z{\'a}r}, J., {Papp}, I., {S{\'a}ri}, P.}, HAT-P-32b and HAT-P-33b: Two Highly Inflated Hot Jupiters Transiting High-jitter Stars, {\it ApJ}, {\bf 742}, 59\new{--77} (2011)

\item Wang, Y.-H., \new{{Wang}, S., {Hinse}, T. C., {Wu}, Z.-Y., {Davis}, A. B., {Hori}, Y., {Yoon}, J.-N., {Han}, W., {Nie}, J.-D., {Liu}, H.-G., {Zhang}, H., {Zhou}, J.-L., {Wittenmyer}, R.~A., {Peng}, X.-Y., {Laughlin}, G.}, Transiting Exoplanet Monitoring Project (TEMP). V. Transit Follow Up for HAT-P-9b, HAT-P-32b, and HAT-P-36b, {\it AJ}, {\bf 157}, 82\new{--98} (2019)

\item Czesla, S., \new{{Lamp{\'o}n}, M., {Sanz-Forcada}, J., {Garc{\'\i}a Mu{\~n}oz}, A., {L{\'o}pez-Puertas}, M., {Nortmann}, L., {Yan}, D., {Nagel}, E., {Yan}, F., {Schmitt}, J.~H.~M.~M., {Aceituno}, J., {Amado}, P.~J., {Caballero}, J.~A., {Casasayas-Barris}, N., {Henning}, Th., {Khalafinejad}, S., {Molaverdikhani}, K., {Montes}, D., {Pall{\'e}}, E., {Reiners}, A., {Schneider}, P.~C., {Ribas}, I., {Quirrenbach}, A., {Zapatero Osorio}, M.~R., {Zechmeister}, M.}, H{\ensuremath{\alpha}} and He I absorption in HAT-P-32 b observed with CARMENES. Detection of Roche lobe overflow and mass loss, {\it A\&A}, {\bf 657}, 6\new{--31} (2022)

\item Quirrenbach, A., \new{{Amado}, P.~J., {Caballero}, J.~A., {Mundt}, R., {Reiners}, A., {Ribas}, I., {Seifert}, W., {Abril}, M., {Aceituno}, J., {Alonso-Floriano}, F.~J., {Ammler-von Eiff}, M., {Antona Jim{\'e}nez}, R., {Anwand-Heerwart}, H., {Azzaro}, M., {Bauer}, F., {Barrado}, D., {Becerril}, S., {B{\'e}jar}, V.~J.~S., {Ben{\'\i}tez}, D., {Berdi{\~n}as}, Z.~M., {C{\'a}rdenas}, M.~C., {Casal}, E., {Claret}, A., {Colom{\'e}}, J., {Cort{\'e}s-Contreras}, M., {Czesla}, S., {Doellinger}, M., {Dreizler}, S., {Feiz}, C., {Fern{\'a}ndez}, M., {Galad{\'\i}}, D., {G{\'a}lvez-Ortiz}, M.~C., {Garc{\'\i}a-Piquer}, A., {Garc{\'\i}a-Vargas}, M.~L., {Garrido}, R., {Gesa}, L., {G{\'o}mez Galera}, V., {Gonz{\'a}lez {\'A}lvarez}, E., {Gonz{\'a}lez Hern{\'a}ndez}, J.~I., {Gr{\"o}zinger}, U., {Gu{\`a}rdia}, J., {Guenther}, E.~W., {de Guindos}, E., {Guti{\'e}rrez-Soto}, J., {Hagen}, H. -J., {Hatzes}, A.~P., {Hauschildt}, P.~H., {Helmling}, J., {Henning}, T., {Hermann}, D., {Hern{\'a}ndez Casta{\~n}o}, L., {Herrero}, E., {Hidalgo}, D., {Holgado}, G., {Huber}, A., {Huber}, K.~F., {Jeffers}, S., {Joergens}, V., {de Juan}, E., {Kehr}, M., {Klein}, R., {K{\"u}rster}, M., {Lamert}, A., {Lalitha}, S., {Laun}, W., {Lemke}, U., {Lenzen}, R., {L{\'o}pez del Fresno}, M., {L{\'o}pez Mart{\'\i}}, B., {L{\'o}pez-Santiago}, J., {Mall}, U., {Mandel}, H., {Mart{\'\i}n}, E.~L., {Mart{\'\i}n-Ruiz}, S., {Mart{\'\i}nez-Rodr{\'\i}guez}, H., {Marvin}, C.~J., {Mathar}, R.~J., {Mirabet}, E., {Montes}, D., {Morales Mu{\~n}oz}, R., {Moya}, A., {Naranjo}, V., {Ofir}, A., {Oreiro}, R., {Pall{\'e}}, E., {Panduro}, J., {Passegger}, V. -M., {P{\'e}rez-Calpena}, A., {P{\'e}rez Medialdea}, D., {Perger}, M., {Pluto}, M., {Ram{\'o}n}, A., {Rebolo}, R., {Redondo}, P., {Reffert}, S., {Reinhardt}, S., {Rhode}, P., {Rix}, H. -W., {Rodler}, F., {Rodr{\'\i}guez}, E., {Rodr{\'\i}guez-L{\'o}pez}, C., {Rodr{\'\i}guez-P{\'e}rez}, E., {Rohloff}, R. -R., {Rosich}, A., {S{\'a}nchez-Blanco}, E., {S{\'a}nchez Carrasco}, M.~A., {Sanz-Forcada}, J., {Sarmiento}, L.~F., {Sch{\"a}fer}, S., {Schiller}, J., {Schmidt}, C., {Schmitt}, J.~H.~M.~M., {Solano}, E., {Stahl}, O., {Storz}, C., {St{\"u}rmer}, J., {Su{\'a}rez}, J.~C., {Ulbrich}, R.~G., {Veredas}, G., {Wagner}, K., {Winkler}, J., {Zapatero Osorio}, M.~R., {Zechmeister}, M., {Abell{\'a}n de Paco}, F.~J., {Anglada-Escud{\'e}}, G., {del Burgo}, C., {Klutsch}, A., {Lizon}, J.~L., {L{\'o}pez-Morales}, M., {Morales}, J.~C., {Perryman}, M.~A.~C., {Tulloch}, S.~M., {Xu}, W.}, CARMENES instrument overview, {\it SPIE}, vol. 9147, p. 91471F, \new{12 pages} (2014)

\item Ninan, J. P., \new{{Bender}, C. F., {Mahadevan}, S., {Ford}, E. B., {Monson}, A. J., {Kaplan}, K. F., {Terrien}, R. C., {Roy}, A., {Robertson}, P. M., {Kanodia}, S., {Stefansson}, G. K.}, The Habitable-Zone Planet Finder: improved flux image generation algorithms for H2RG up-the-ramp data, {\it SPIE}, vol. 10709, p. 107092U, \new{11 pages} (2018)

\item Kaplan, K. F., \new{{Bender}, C. F., {Terrien}, R. C., {Ninan}, J., {Roy}, A., {Mahadevan}, S.}, The algorithms behind the HPF and NEID pipeline, {\it ASPCS}, {\bf 523}, 567\new{--570} (2019)

\item Metcalf, A. J., \new{{Anderson}, T., {Bender}, C. F., {Blakeslee}, S., {Brand}, W., {Carlson}, D. R., {Cochran}, W. D., {Diddams}, S. A., {Endl}, M., {Fredrick}, C., {Halverson}, S., {Hickstein}, D. D., {Hearty}, F., {Jennings}, J., {Kanodia}, S., {Kaplan}, K. F., {Levi}, E., {Lubar}, E., {Mahadevan}, S., {Monson}, A., {Ninan}, J. P., {Nitroy}, C., {Osterman}, S., {Papp}, S. B., {Quinlan}, F., {Ramsey}, L., {Robertson}, P., {Roy}, A., {Schwab}, C., {Sigurdsson}, S., {Srinivasan}, K., {Stefansson}, G., {Sterner}, D. A., {Terrien}, R., {Wolszczan}, A., {Wright}, J. T., {Ycas}, G.}, Stellar spectroscopy in the near-infrared with a laser frequency comb, {\it Optica}, {\bf 6}, 233\new{--239} (2019)

\item Gully-Santiago, M. \new{{Luna}, J., {Morley}, C., {Kaplan}, K., {Ganesh}, A., {Sawczynec}, E., {Burke}, J., {Krolikowski}, D.}, Astronomical {\'e}chelle spectroscopy data analysis with \texttt{muler}, The Journal of Open Source Software, {\bf 7}, 4302 (2022)

\item Owen, J. \new{and {Adams}, F. C.}, Magnetically controlled mass-loss from extrasolar planets in close orbits, {\it MNRAS}, {\bf 444}, 3761\new{--3779} (2014)

\item Spake, J. J. \new{{Oklop{\v{c}}i{\'c}}, A., {Hillenbrand}, L.~A.}, The Posttransit Tail of WASP-107b Observed at 10830 {\r{A}}, {\it AJ}, {\bf 162}, 284\new{--292} (2021)

\item Ehrenreich, D., \new{{Bourrier}, V., {Wheatley}, P. J., {Lecavelier des Etangs}, A., {H{\'e}brard}, G., {Udry}, S., {Bonfils}, X., {Delfosse}, X., {D{\'e}sert}, J.-M., {Sing}, D. K., {Vidal-Madjar}, A.}, A giant comet-like cloud of hydrogen escaping the warm Neptune-mass exoplanet GJ 436b, {\it Nature}, {\bf 522}, 459\new{--461} (2015)

\item Lavie, B., \new{{Ehrenreich}, D., {Bourrier}, V., {Lecavelier des Etangs}, A., {Vidal-Madjar}, A., {Delfosse}, X., {Gracia Berna}, A., {Heng}, K., {Thomas}, N., {Udry}, S., {Wheatley}, P.~J.}, The long egress of GJ 436b's giant exosphere, {\it A\&A}, {\bf 605}, 7\new{--13} (2017)

\item MacLeod, M. \new{and {Oklop{\v{c}}i{\'c}}, A.}, Stellar Wind Confinement of Evaporating Exoplanet Atmospheres and Its Signatures in 1083 nm Observations, {\it ApJ}, {\bf 926}, 226\new{--238} (2022)

\item Oklop{\v{c}}i{\'c}, A., Helium Absorption at 1083 nm from Extended Exoplanet Atmospheres: Dependence on Stellar Radiation, {\it ApJ}, {\bf 881}, 133\new{--140} (2019)

\item {Carleo}, I., \new{{Youngblood}, A., {Redfield}, S., {Casasayas Barris}, N., {Ayres}, T. R., {Vannier}, H., {Fossati}, L., {Palle}, E., {Livingston}, J. H., {Lanza}, A. F., {Niraula}, P., {Alvarado-G{\'o}mez}, J., {Chen}, G., {Gandolfi}, D., {Guenther}, E. W., {Linsky}, J. L., {Nagel}, E., {Narita}, N., {Nortmann}, L., {Shkolnik}, E.  L.,  {Stangret}, M.}, A Multiwavelength Look at the GJ 9827 System: No Evidence of Extended Atmospheres in GJ 9827b and d from HST and CARMENES Data, {\it AJ}, {\bf 161}, 136\new{--147} (2021)

\item Zhang, M., \new{{Knutson}, H. A., {Wang}, L., {Dai}, F., {Barrag{\'a}n}, O.}, Escaping Helium from TOI 560.01, a Young Mini-Neptune, {\it AJ}, {\bf 163}, 67\new{--81} (2022)

\item {H{\'e}brard}, G., \new{{Collier Cameron}, A., {Brown}, D.~J.~A., {D{\'\i}az}, R.~F., {Faedi}, F., {Smalley}, B., {Anderson}, D.~R., {Armstrong}, D., {Barros}, S.~C.~C., {Bento}, J., {Bouchy}, F., {Doyle}, A.~P., {Enoch}, B., {G{\'o}mez Maqueo Chew}, Y., {H{\'e}brard}, {\'E}. M., {Hellier}, C., {Lendl}, M., {Lister}, T.~A., {Maxted}, P.~F.~L., {McCormac}, J., {Moutou}, C., {Pollacco}, D., {Queloz}, D., {Santerne}, A., {Skillen}, I., {Southworth}, J., {Tregloan-Reed}, J., {Triaud}, A.~H.~M.~J., {Udry}, S., {Vanhuysse}, M., {Watson}, C.~A., {West}, R.~G., {Wheatley}, P.~J.}, WASP-52b, WASP-58b, WASP-59b, and WASP-60b: Four new transiting close-in giant planets, {\it A\&A}, {\bf 549}, 134\new{--144} (2013)

\item Zhang, M. \new{{Knutson}, H. A., {Dai}, F., {Wang}, L., {Ricker}, G. R., {Schwarz}, R. P., {Mann}, C., {Collins}, K.}, Detection of Atmospheric Escape from Four Young Mini Neptunes, {\it AJ}, \new{{\bf 165}, 62--77} (2023)

\item \new{Gaidos, E., {Hirano}, T., {Lee}, R. A., {Harakawa}, H., {Hodapp}, K., {Jacobson}, S., {Kotani}, T., {Kudo}, T., {Kurokawa}, T. {Kuzuhara}, M., {Nishikawa}, J., {Omiya}, M., {Serizawa}, T., {Tamura}, M., {Ueda}, A., {Vievard}, S., Planet(esimal)s around stars with TESS (PAST) III: A search for triplet He I in the atmospheres of two 200 Myr-old planets, {\it MNRAS}, {\bf 518}, 3777--3783 (2023)}

\item Eggleton, P.. P. Aproximations to the radii of Roche lobes, {\it ApJ}, {\bf 268}, 368\new{--369} (1983)

\item Paragas, K., \new{{Vissapragada}, S., {Knutson}, H. A., {Oklop{\v{c}}i{\'c}}, A., {Chachan}, Y., {Greklek-McKeon}, M., {Dai}, F., {Tinyanont}, S., {Vasisht}, G.}, Metastable Helium Reveals an Extended Atmosphere for the Gas Giant HAT-P-18b, {\it ApJ}, {\bf 909}, 10\new{--16} (2021)

\item Bailer-Jones, C. A. L., \new{{Rybizki}, J., {Fouesneau}, M., {Demleitner}, M., {Andrae}, R.} Estimating Distances from Parallaxes. V. Geometric and Photogeometric Distances to 1.47 Billion Stars in Gaia Early Data Release 3, {\it AJ}, {\bf 161}, 147\new{--170} (2021)

\item Brown, D. J. A., Discrepancies between isochrone fitting and gyrochronology for exoplanet host stars?, {\it MNRAS}, {\bf 442}, 1844\new{--1862} (2014)

\item Knutson, H. A., \new{{Fulton}, B. J., {Montet}, B. T., {Kao}, M., {Ngo}, H., {Howard}, A. W., {Crepp}, J. R., {Hinkley}, S., {Bakos}, G. {\'A}., {Batygin}, K., {Johnson}, J. A., {Morton}, T. D., {Muirhead}, P. S.}, Friends of Hot Jupiters. I. A Radial Velocity Search for Massive, Long-Period Companions to Close-in Gas Giant Planets, {\it ApJ}, {\bf 785}, 126\new{--148} (2014)

\item Zhao, M., \new{{O'Rourke}, J. G., {Wright}, J. T., {Knutson}, H. A., {Burrows}, A., {Fortney}, J., {Ngo}, H., {Fulton}, B. J., {Baranec}, C., {Riddle}, R., {Law}, N. M., {Muirhead}, P. S., {Hinkley}, S., {Showman}, A. P., {Curtis}, J., {Burruss}, R.}, Characterization of the Atmosphere of the Hot Jupiter HAT-P-32Ab and the M-dwarf Companion HAT-P-32B, {\it ApJ}, {\bf 796}, 115\new{--129} (2014)

\item \new{Fortney, J. J., \new{{Marley}, M.~S., {Barnes}, J.~W.}, Planetary Radii across Five Orders of Magnitude in Mass and Stellar Insolation: Application to Transits, {\it ApJ}, {\bf 659}, 1661\new{--1672} (2007)}

\item Albrecht, S., \new{{Winn}, J. N., {Johnson}, J. A., {Howard}, A. W., {Marcy}, G. W., {Butler}, R. P., {Arriagada}, P., {Crane}, J. D., {Shectman}, S. A., {Thompson}, I. B., {Hirano}, T., {Bakos}, G., {Hartman}, J. D.}, Obliquities of Hot Jupiter Host Stars: Evidence for Tidal Interactions and Primordial Misalignments, {\it AJ}, {\bf 757}, 18\new{--42} (2012)

\item Adams, E. R., \new{{Dupree}, A.~K., {Kulesa}, C., {McCarthy}, D.}, Adaptive Optics Images. II. 12 Kepler Objects of Interest and 15 Confirmed Transiting Planets, {\it AJ}, {\bf 146}, 9\new{13} (2013)

\item Ngo, H., \new{{Knutson}, H. A., {Hinkley}, S., {Crepp}, J. R., {Bechter}, E. B., {Batygin}, K., {Howard}, A. W., {Johnson}, J. A., {Morton}, T. D., {Muirhead}, P. S.}, Friends of Hot Jupiters. II. No Correspondence between Hot-jupiter Spin-Orbit Misalignment and the Incidence of Directly Imaged Stellar Companions, {\it ApJ}, {\bf 800}, 138\new{--159} (2015)

\item \new{Krolikowski, D. {\it et al.}, Exploring the formation and evolutionary pathways of young stars and planetary systems at high precision, {\it Ph.D. Thesis}, University of Texas at Austin (2022), \url{https://repositories.lib.utexas.edu/handle/2152/116932}  }

\item Oliva, E. \new{{Origlia}, L., {Scuderi}, S., {Benatti}, S., {Carleo}, I., {Lapenna}, E., {Mucciarelli}, A., {Baffa}, C., {Biliotti}, V., {Carbonaro}, L., {Falcini}, G., {Giani}, E., {Iuzzolino}, M., {Massi}, F., {Sanna}, N., {Sozzi}, M., {Tozzi}, A., {Ghedina}, A., {Ghinassi}, F., {Lodi}, M., {Harutyunyan}, A., {Pedani}, M.}, Lines and continuum sky emission in the near infrared: observational constraints from deep high spectral resolution spectra with GIANO-TNG, {\it A\&A}, {\bf 581}, 47\new{--52} (2015)

\item Tran, Q. H., \new{{Tran}, Q. H., {Bowler}, B. P., {Cochran}, W. D., {Endl}, M., {Stef{\'a}nsson}, G., {Mahadevan}, S., {Ninan}, J. P., {Bender}, C. F., {Halverson}, S., {Roy}, A., and {Terrien}, R. C.}, The Epoch of Giant Planet Migration Planet Search Program. I. Near-infrared Radial Velocity Jitter of Young Sun-like Stars, {\it AJ}, {\bf 161}, 173\new{189} (2021)

\item Astropy Collaboration, Robitaille, T. P., \new{{Tollerud}, E. J., {Greenfield}, P., {Droettboom}, M., {Bray}, E., {Aldcroft}, T., {Davis}, M., {Ginsburg}, A., {Price-Whelan}, A. M., {Kerzendorf}, W. E., {Conley}, A., {Crighton}, N., {Barbary}, K., {Muna}, D., {Ferguson}, H., {Grollier}, F., {Parikh}, M. M., {Nair}, P. H., {Unther}, H. M., {Deil}, C., {Woillez}, J., {Conseil}, S., {Kramer}, R., {Turner}, J. E.~H., {Singer}, L., {Fox}, R., {Weaver}, B. A., {Zabalza}, V., {Edwards}, Z. I., {Azalee Bostroem}, K., {Burke}, D.~J., {Casey}, A. R., {Crawford}, S. M., {Dencheva}, N., {Ely}, J., {Jenness}, T., {Labrie}, K., {Lim}, P. L., {Pierfederici}, F., {Pontzen}, A., {Ptak}, A., {Refsdal}, B., {Servillat}, M., {Streicher}, O.}, Astropy: A community Python package for astronomy, {\it A\&A}, {\bf 558}, 33\new{--41} (2013)

\item Astropy Collaboration, Price-Whelan, A. M., \new{{Sip{\H{o}}cz}, B.~M., {G{\"u}nther}, H.~M., {Lim}, P.~L., {Crawford}, S.~M., {Conseil}, S., {Shupe}, D.~L., {Craig}, M.~W., {Dencheva}, N., {Ginsburg}, A., {V,erPlas}, J.~T., {Bradley}, L.~D., {P{\'e}rez-Su{\'a}rez}, D., {de Val-Borro}, M., {Aldcroft}, T.~L., {Cruz}, K.~L., {Robitaille}, T.~P., {Tollerud}, E.~J., {Ardelean}, C., {Babej}, T., {Bach}, Y.~P., {Bachetti}, M., {Bakanov}, A.~V., {Bamford}, S.~P., {Barentsen}, G., {Barmby}, P., {Baumbach}, A., {Berry}, K.~L., {Biscani}, F., {Boquien}, M., {Bostroem}, K.~A., {Bouma}, L.~G. , {Brammer}, G.~B. , {Bray}, E.~M. , {Breytenbach}, H. , {Buddelmeijer}, H. , {Burke}, D.~J. , {Calderone}, G. , {Cano Rodr{\'\i}guez}, J.~L. , {Cara}, M. , {Cardoso}, J.~V.~M. , {Cheedella}, S. , {Copin}, Y. , {Corrales}, L. , {Crichton}, D. , {D'Avella}, D. , {Deil}, C. , {Depagne}, {\'E}. , {Dietrich}, J.~P. , {Donath}, A. , {Droettboom}, M. , {Earl}, N. , {Erben}, T. , {Fabbro}, S. , {Ferreira}, L.~A. , {Finethy}, T. , {Fox}, R.~T. , {Garrison}, L.~H. , {Gibbons}, S.~L.~J. , {Goldstein}, D.~A. , {Gommers}, R. , {Greco}, J.~P. , {Greenfield}, P. , {Groener}, A.~M. , {Grollier}, F. , {Hagen}, A. , {Hirst}, P. , {Homeier}, D. , {Horton}, A.~J. , {Hosseinzadeh}, G. , {Hu}, L. , {Hunkeler}, J.~S. , {Ivezi{\'c}}, {\v{Z}}. , {Jain}, A. , {Jenness}, T. , {Kanarek}, G. , {Kendrew}, S. , {Kern}, N.~S. , {Kerzendorf}, W.~E. , {Khvalko}, A. , {King}, J. , {Kirkby}, D. , {Kulkarni}, A.~M. , {Kumar}, A. , {Lee}, A. , {Lenz}, D. , {Littlefair}, S.~P. , {Ma}, Z. , {Macleod}, D.~M. , {Mastropietro}, M. , {McCully}, C. , {Montagnac}, S. , {Morris}, B.~M. , {Mueller}, M. , {Mumford}, S.~J. , {Muna}, D. , {Murphy}, N.~A. , {Nelson}, S. , {Nguyen}, G.~H. , {Ninan}, J.~P. , {N{\"o}the}, M. , {Ogaz}, S. , {Oh}, S. , {Parejko}, J.~K. , {Parley}, N. , {Pascual}, S. , {Patil}, R. , {Patil}, A.~A. , {Plunkett}, A.~L. , {Prochaska}, J.~X. , {Rastogi}, T. , {Reddy Janga}, V. , {Sabater}, J. , {Sakurikar}, P. , {Seifert}, M. , {Sherbert}, L.~E. , {Sherwood-Taylor}, H. , {Shih}, A.~Y. , {Sick}, J. , {Silbiger}, M.~T. , {Singanamalla}, S. , {Singer}, L.~P. , {Sladen}, P.~H. , {Sooley}, K.~A. , {Sornarajah}, S. , {Streicher}, O. , {Teuben}, P. , {Thomas}, S.~W. , {Tremblay}, G.~R. , {Turner}, J.~E.~H. , {Terr{\'o}n}, V. , {van Kerkwijk}, M.~H. , {de la Vega}, A. , {Watkins}, L.~L. , {Weaver}, B.~A. , {Whitmore}, J.~B. , {Woillez}, J. , {Zabalza}, V. , {Astropy Contributors}}, The Astropy Project: Building an Open-science Project and Status of the v2.0 Core Package, {\it AJ}, {\bf 156}, 123\new{--141} (2018)

\item Zechmeister, M., \new{{Reiners}, A., {Amado}, P.~J., {Azzaro}, M., {Bauer}, F.~F., {B{\'e}jar}, V.~J.~S., {Caballero}, J.~A., {Guenther}, E.~W., {Hagen}, H. -J., {Jeffers}, S.~V., {Kaminski}, A., {K{\"u}rster}, M., {Launhardt}, R., {Montes}, D., {Morales}, J.~C., {Quirrenbach}, A., {Reffert}, S., {Ribas}, I., {Seifert}, W., {Tal-Or}, L., {Wolthoff}, V.}, Spectrum radial velocity analyser (SERVAL). High-precision radial velocities and two alternative spectral indicators, {\it A\&A}, {\bf 609}, 12\new{--24} (2018)

\item Martin, J., \new{{Fuhrmeister}, B., {Mittag}, M., {Schmidt}, T.~O.~B., {Hempelmann}, A., {Gonz{\'a}lez-P{\'e}rez}, J.~N., {Schmitt}, J.~H.~M.~M.}, The Ca II infrared triplet's performance as an activity indicator compared to Ca II H and K. Empirical relations to convert Ca II infrared triplet measurements to common activity indices, {\it A\&A}, {\bf 605}, 113\new{--127} (2017)

\item Alam, M. K., \new{{L{\'o}pez-Morales}, M., {Nikolov}, N., {Sing}, D. K., {Henry}, G. W., {Baxter}, C., {D{\'e}sert}, J.-M., {Barstow}, J. K., {Mikal-Evans}, T., {Bourrier}, V., {Lavvas}, P., {Wakeford}, H. R., {Williamson}, M. H., {Sanz-Forcada}, J., {Buchhave}, L. A., {Cohen}, O., {Garc{\'\i}a Mu{\~n}oz}, A.}, The Hubble Space Telescope PanCET Program: An Optical to Infrared Transmission Spectrum of HAT-P-32Ab, {\it AJ}, {\bf 160}, 51\new{--69} (2020)

\item Stone, J. M., \new{{Tomida}, K., {White}, C. J., {Felker}, K. G.}, The Athena++ Adaptive Mesh Refinement Framework: Design and Magnetohydrodynamic Solvers, {\it ApJS}, {\bf 249}, 4\new{--43} (2020)

\item Linssen, D., \new{{Oklop{\v{c}}i{\'c}}, A., {MacLeod}, M.}, Constraining planetary mass-loss rates by simulating Parker wind profiles with Cloudy, {\it A\&A}, \new{{\bf 667}, 54--67} (2022)

\item Wood, B. E., \new{ {M{\"u}ller}, H.-R., {Zank}, G. P., {Linsky}, J. L.}, Measured Mass-Loss Rates of Solar-like Stars as a Function of Age and Activity, {\it ApJ}, {\bf 574}, 412\new{--425} (2002)

\item \new{Sanz-Forcada, J. \new{{Micela}, G., {Ribas}, I., {Pollock}, A.~M.~T., {Eiroa}, C., {Velasco}, A., {Solano}, E., {Garc{\'\i}a-{\'A}lvarez}, D.}, Estimation of the XUV radiation onto close planets and their evaporation, {\it A\&A}, {\bf 532}, 6\new{--23} (2011)}

\item \new{Oklop{\v{c}}i{\'c}, A., \new{and} Hirata, C.M., A New Window into Escaping Exoplanet Atmospheres: 10830~\AA\ Line of Helium, {\it ApJ}, {\bf 855}, 11\new{--17} (2018)}

\item Ninan, J. P., \new{{Stefansson}, G., {Mahadevan}, S., {Bender}, C., {Robertson}, P., {Ramsey}, L., {Terrien}, R., {Wright}, J., {Diddams}, S. A., {Kanodia}, S., {Cochran}, W., {Endl}, M., {Ford}, E. B., {Fredrick}, C., {Halverson}, S., {Hearty}, F., {Jennings}, J., {Kaplan}, K., {Lubar}, E., {Metcalf}, A. J., {Monson}, A., {Nitroy}, C., {Roy}, A., {Schwab}, C.}, Evidence for He I 10830 {\r{A}} Absorption during the Transit of a Warm Neptune around the M-dwarf GJ 3470 with the Habitable-zone Planet Finder, {\it ApJ}, {\bf 894}, 97\new{--105} (2020)

\item Mann, A. W., \new{{Vanderburg}, A., {Rizzuto}, A. C., {Kraus}, A. L., {Berlind}, P., {Bieryla}, A., {Calkins}, M. L., {Esquerdo}, G. A., {Latham}, D. W., {Mace}, G. N., {Morris}, N. R., {Quinn}, S. N., {Sokal}, K. R., {Stefanik}, R. P.}, Zodiacal Exoplanets in Time (ZEIT). VI. A Three-planet System in the Hyades Cluster Including an Earth-sized Planet, {\it AJ}, {\bf 155}, 4\new{--14} (2018)

\item Ciardi, D. R., \new{{Crossfield}, I. J.~M., {Feinstein}, A. D., {Schlieder}, J. E., {Petigura}, E. A., {David}, T. J., {Bristow}, M., {Patel}, R. I., {Arnold}, L., {Benneke}, B., {Christiansen}, J. L., {Dressing}, C. D., {Fulton}, B. J., {Howard}, A. W., {Isaacson}, H., {Sinukoff}, E., {Thackeray}, B.}, K2-136: A Binary System in the Hyades Cluster Hosting a Neptune-sized Planet, {\it AJ}, {\bf 155}, 10\new{--20} (2018)

\item Gaidos, E., \new{Hirano, T., Omiya, M., Kuzuhara, M., Kotani, T., Tamura, M., Harakawa, H., Kudo, T.}, Zodiacal Exoplanets in Time (ZEIT). XIV. He I Transit Spectroscopy of the 650 Myr Hyades Planet K2-136c, {\it RNAAS}, {\bf 5}, 238 (2021)

\item Fern{\'a}ndez, J., \new{and} Wheatley, P. J., X-ray irradiation of three planets around Hyades star K2 -136, {\it AN}, 34310076, \new{7 pages} (2022)

\item {Barrag{\'a}n}, O., \new{{Aigrain}, S., {Kubyshkina}, D., {Gandolfi}, D., {Livingston}, J., {Fridlund}, M.~C.~V., {Fossati}, L., {Korth}, J., {Parviainen}, H., {Malavolta}, L., {Palle}, E., {Deeg}, H.~J., {Nowak}, G., {Rajpaul}, V.~M., {Zicher}, N., {Antoniciello}, G., {Narita}, N., {Albrecht}, S., {Bedin}, L.~R., {Cabrera}, J., {Cochran}, W.~D., {de Leon}, J., {Eigm{\"u}ller}, Ph, {Fukui}, A., {Granata}, V., {Grziwa}, S., {Guenther}, E., {Hatzes}, A.~P., {Kusakabe}, N., {Latham}, D.~W., {Libralato}, M., {Luque}, R., {Monta{\~n}{\'e}s-Rodr{\'\i}guez}, P., {Murgas}, F., {Nardiello}, D., {Pagano}, I., {Piotto}, G., {Persson}, C.~M., {Redfield}, S., {Tamura}, M.}, Radial velocity confirmation of K2-100b: a young, highly irradiated, and low-density transiting hot Neptune, {\it MNRAS}, {\bf 490}, 698\new{--708} (2019)

\item Gaidos, E., \new{{Hirano}, T., {Mann}, A.~W., {Owens}, D.~A., {Berger}, T.~A., {France}, K., {Vanderburg}, A., {Harakawa}, H., {Hodapp}, K.~W., {Ishizuka}, M., {Jacobson}, S., {Konishi}, M., {Kotani}, T., {Kudo}, T., {Kurokawa}, T., {Kuzuhara}, M., {Nishikawa}, J., {Omiya}, M., {Serizawa}, T., {Tamura}, M., {Ueda}, A.}, Zodiacal exoplanets in time - X. The orbit and atmosphere of the young 'neptune desert'-dwelling planet K2-100b, {\it MNRAS}, {\bf 495}, 650\new{--662} (2020)

\item David, T. J., \new{{Petigura}, E. A., {Luger}, R., {Foreman-Mackey}, D., {Livingston}, J. H., {Mamajek}, E. E., {Hillenbrand}, L. A.}, Four Newborn Planets Transiting the Young Solar Analog V1298 Tau, {\it ApJ}, {\bf 885}, 12\new{--21} (2019)

\item Vissapragada, S., \new{{Stef{\'a}nsson}, G., {Greklek-McKeon}, M., {Oklop{\v{c}}i{\'c}}, A., {Knutson}, H. A., {Ninan}, J. P., {Mahadevan}, S., {Ca{\~n}as}, C. I., {Chachan}, Y., {Cochran}, W. D., {Collins}, K. A., {Dai}, F., {David}, T. J., {Halverson}, S., {Hawley}, S. L., {Hebb}, L., {Kanodia}, S., {Kowalski}, A. F., {Livingston}, J. H., {Maney}, M., {Metcalf}, A. J., {Morley}, C., {Ramsey}, L. W., {Robertson}, P., {Roy}, A., {Spake}, J., {Schwab}, C., {Terrien}, R. C., {Tinyanont}, S., {Vasisht}, G., {Wisniewski}, J.}, A Search for Planetary Metastable Helium Absorption in the V1298 Tau System, {\it AJ}, {\bf 162}, 222\new{--231} (2021)

\item {Poppenhaeger}, K., \new{{Ketzer}, L., {Mallonn}, M.}, X-ray irradiation and evaporation of the four young planets around V1298 Tau, {\it MNRAS}, {\bf 500}, 4560\new{--4572} (2021)

\item {Gillon}, M., \new{{Triaud}, A. H.~M.~J., {Demory}, B.-O., {Jehin}, E., {Agol}, E., {Deck}, K. M., {Lederer}, S. M., {de Wit}, J., {Burdanov}, A., {Ingalls}, J. G., {Bolmont}, E., {Leconte}, J., {Raymond}, S. N., {Selsis}, F., {Turbet}, M., {Barkaoui}, K., {Burgasser}, A., {Burleigh}, M. R., {Carey}, S. J., {Chaushev}, A., {Copperwheat}, C. M., {Delrez}, L., {Fernandes}, C. S., {Holdsworth}, D. L., {Kotze}, E. J., {Van Grootel}, V., {Almleaky}, Y., {Benkhaldoun}, Z., {Magain}, P., {Queloz}, D.}, Seven temperate terrestrial planets around the nearby ultracool dwarf star TRAPPIST-1, {\it Nature}, {\bf 542}, 456\new{--460} (2017)

\item {Wheatley}, P. J., \new{{Louden}, T., {Bourrier}, V., {Ehrenreich}, D., {Gillon}, M.}, Strong XUV irradiation of the Earth-sized exoplanets orbiting the ultracool dwarf TRAPPIST-1, {\it MNRAS}, {\bf 465}, 74\new{--78} (2017)

\item {Krishnamurthy}, V., \new{{Hirano}, T., {Stef{\'a}nsson}, G., {Ninan}, J. P., {Mahadevan}, S., {Gaidos}, E., {Kopparapu}, R., {Sato}, B., {Hori}, Y., {Bender}, C. F., {Ca{\~n}as}, C. I., {Diddams}, S. A., {Halverson}, S., {Harakawa}, H., {Hawley}, S., {Hearty}, F., {Hebb}, L., {Hodapp}, K., {Jacobson}, S., {Kanodia}, S., {Konishi}, M., {Kotani}, T., {Kowalski}, A., {Kudo}, T., {Kurokawa}, T., {Kuzuhara}, M., {Lin}, A., {Maney}, M., {Metcalf}, A. J., {Morris}, B., {Nishikawa}, J., {Omiya}, M., {Robertson}, P., {Roy}, A., {Schwab}, C., {Serizawa}, T., {Tamura}, M., {Ueda}, A., {Vievard}, S., {Wisniewski}, J.}, Nondetection of Helium in the Upper Atmospheres of TRAPPIST-1b, e, and f, {\it AJ}, {\bf 162}, 82\new{--89} (2021)

\item {Agol}, E., \new{{Dorn}, C., {Grimm}, S. L., {Turbet}, M., {Ducrot}, E., {Delrez}, L., {Gillon}, M., {Demory}, B.-O., {Burdanov}, A., {Barkaoui}, K., {Benkhaldoun}, Z., {Bolmont}, E., {Burgasser}, A., {Carey}, S., {de Wit}, J., {Fabrycky}, D., {Foreman-Mackey}, D., {Haldemann}, J., {Hernandez}, D. M., {Ingalls}, J., {Jehin}, E., {Langford}, Z., {Leconte}, J., {Lederer}, S. M., {Luger}, R., {Malhotra}, R., {Meadows}, V. S., {Morris}, B. M., {Pozuelos}, F. J., {Queloz}, D., {Raymond}, S. N., {Selsis}, F., {Sestovic}, M., {Triaud}, A. H.~M.~J., {Van Grootel}, V.}, Refining the Transit-timing and Photometric Analysis of TRAPPIST-1: Masses, Radii, Densities, Dynamics, and Ephemerides, {\it PSJ}, {\bf 2}, 1\new{--38} (2021)

\item {Turner}, O. D., \new{{Anderson}, D.~R., {Barkaoui}, K., {Bouchy}, F., {Benkhaldoun}, Z., {Brown}, D.~J.~A., {Burdanov}, A., {Collier Cameron}, A., {Ducrot}, E., {Gillon}, M., {Hellier}, C., {Jehin}, E., {Lendl}, M., {Maxted}, P.~F.~L., {Nielsen}, L.~D., {Pepe}, F., {Pollacco}, D., {Pozuelos}, F.~J., {Queloz}, D., {S{\'e}gransan}, D., {Smalley}, B., {Triaud}, A.~H.~M.~J., {Udry}, S., {West}, R.~G.}, Three hot-Jupiters on the upper edge of the mass-radius distribution: WASP-177, WASP-181, and WASP-183, {\it MNRAS}, {\bf 485}, 5790\new{--5799} (2019)

\item {Kirk}, J., \new{{Dos Santos}, L. A., {L{\'o}pez-Morales}, M., {Alam}, M. K., {Oklop{\v{c}}i{\'c}}, A., {MacLeod}, M., {Zeng}, L., {Zhou}, G.}, Keck/NIRSPEC Studies of He I in the Atmospheres of Two Inflated Hot Gas Giants Orbiting K Dwarfs: WASP-52b and WASP-177b, {\it AJ}, {\bf 164}, 24\new{--37} (2022)

\item {Kasper}, D., \new{{Bean}, J. L., {Oklop{\v{c}}i{\'c}}, A., {Malsky}, I., {Kempton}, E. M. -R., {D{\'e}sert}, J.-M., {Rogers}, L. A., {Mansfield}, M.}, Nondetection of Helium in the Upper Atmospheres of Three Sub-Neptune Exoplanets, {\it AJ}, {\bf 160}, 258\new{--270} (2020)

\item {Rice}, K., \new{{Malavolta}, L., {Mayo}, A., {Mortier}, A., {Buchhave}, L.~A., {Affer}, L., {Vanderburg}, A., {Lopez-Morales}, M., {Poretti}, E., {Zeng}, L., {Collier Cameron}, A., {Damasso}, M., {Coffinet}, A., {Latham}, D.~W., {Bonomo}, A.~S., {Bouchy}, F., {Charbonneau}, D., {Dumusque}, X., {Figueira}, P., {Martinez Fiorenzano}, A.~F., {Haywood}, R.~D., {Johnson}, J. A., {Lopez}, E., {Lovis}, C., {Mayor}, M., {Micela}, G., {Molinari}, E., {Nascimbeni}, V., {Nava}, C., {Pepe}, F., {Phillips}, D.~F., {Piotto}, G., {Sasselov}, D., {S{\'e}gransan}, D., {Sozzetti}, A., {Udry}, S., {Watson}, C.}, Masses and radii for the three super-Earths orbiting GJ 9827, and implications for the composition of small exoplanets, {\it MNRAS}, {\bf 484}, 3731\new{--3745} (2019)

\item {Ellis}, T. G., \new{{Boyajian}, T., {von Braun}, K., {Ligi}, R., {Mourard}, D., {Dragomir}, D., {Schaefer}, G. H., {Farrington}, C. D.}, Directly Determined Properties of HD 97658 from Interferometric Observations, {\it AJ}, {\bf 162}, 118\new{--126} (2021)

\item {Demory}, B.-O., \new{{Gillon}, M., {Deming}, D., {Valencia}, D., {Seager}, S., {Benneke}, B., {Lovis}, C., {Cubillos}, P., {Harrington}, J., {Stevenson}, K.~B., {Mayor}, M., {Pepe}, F., {Queloz}, D., {S{\'e}gransan}, D., {Udry}, S.}, Detection of a transit of the super-Earth 55 Cancri e with warm Spitzer, {\it A\&A}, {\bf 533}, 114\new{--120} (2011)

\item {Bourrier}, V., \new{{Dumusque}, X., {Dorn}, C., {Henry}, G.~W., {Astudillo-Defru}, N., {Rey}, J., {Benneke}, B., {H{\'e}brard}, G., {Lovis}, C., {Demory}, B.~O., {Moutou}, C., {Ehrenreich}, D.}, The 55 Cancri system reassessed, {\it A\&A}, {\bf 619}, 1\new{--18} (2018)

\item {Zhang}, M., \new{{Knutson}, H. A., {Wang}, L., {Dai}, F., {Oklopcic}, A., {Hu}, R.}, No Escaping Helium from 55 Cnc e, {\it AJ}, {\bf 161}, 181\new{--198} (2021)

\item {Gaudi}, B.S., \new{{Stassun}, K. G., {Collins}, K. A., {Beatty}, T. G., {Zhou}, G., {Latham}, D. W., {Bieryla}, A., {Eastman}, J. D., {Siverd}, R. J., {Crepp}, J. R., {Gonzales}, E. J., {Stevens}, D. J., {Buchhave}, L. A., {Pepper}, J., {Johnson}, M. C., {Colon}, K. D., {Jensen}, E. L.~N., {Rodriguez}, J. E., {Bozza}, V., {Novati}, S. C., {D'Ago}, G., {Dumont}, M. T., {Ellis}, T., {Gaillard}, C., {Jang-Condell}, H., {Kasper}, D. H., {Fukui}, A., {Gregorio}, J., {Ito}, A., {Kielkopf}, J. F., {Manner}, M., {Matt}, K., {Narita}, N., {Oberst}, T. E., {Reed}, P. A., {Scarpetta}, G., {Stephens}, D. C., {Yeigh}, R. R., {Zambelli}, R., {Fulton}, B.~J., {Howard}, A. W., {James}, D. J., {Penny}, M., {Bayliss}, D., {Curtis}, I. A., {Depoy}, D.~L., {Esquerdo}, G. A., {Gould}, A., {Joner}, M. D., {Kuhn}, R. B., {Labadie-Bartz}, J., {Lund}, M. B., {Marshall}, J. L., {McLeod}, K. K., {Pogge}, R. W., {Relles}, H., {Stockdale}, C., {Tan}, T.~G., {Trueblood}, M., {Trueblood}, P.}, A giant planet undergoing extreme-ultraviolet irradiation by its hot massive-star host, {\it Nature}, {\bf 546}, 514\new{--518} (2017)

\item {Torres}, G., \new{ {Winn}, J. N., {Holman}, M. J.}, Improved Parameters for Extrasolar Transiting Planets, {\it ApJ}, {\bf 677}, 1324\new{--1342} (2008)

\item {Maciejewski}, G., \new{{Niedzielski}, A., {Nowak}, G., {Pall{\'e}}, E., {Tingley}, B., {Errmann}, R., {Neuh{\"a}user}, R.}, On the GJ 436 Planetary System, {\it AcA}, {\bf 64}, 323\new{--335} (2014)

\item {Triaud}, A.H.M.J., \new{{Gillon}, M., {Ehrenreich}, D., {Herrero}, E., {Lendl}, M., {Anderson}, D. R., {Collier Cameron}, A., {Delrez}, L., {Demory}, B.-O., {Hellier}, C., {Heng}, K., {Jehin}, E., {Maxted}, P. F.~L., {Pollacco}, D., {Queloz}, D., {Ribas}, I., {Smalley}, B., {Smith}, A. M.~S., {Udry}, S.}, WASP-80b has a dayside within the T-dwarf range, {\it MNRAS}, {\bf 450}, 2279\new{--2290} (2015)

\item {Fossati}, L., \new{{Guilluy}, G., {Shaikhislamov}, I.~F., {Carleo}, I., {Borsa}, F., {Bonomo}, A.~S., {Giacobbe}, P., {Rainer}, M., {Cecchi-Pestellini}, C., {Khodachenko}, M.~L., {Efimov}, M.~A., {Rumenskikh}, M.~S., {Miroshnichenko}, I.~B., {Berezutsky}, A.~G., {Nascimbeni}, V., {Brogi}, M., {Lanza}, A.~F., {Mancini}, L., {Affer}, L., {Benatti}, S., {Biazzo}, K., {Bignamini}, A., {Carosati}, D., {Claudi}, R., {Cosentino}, R., {Covino}, E., {Desidera}, S., {Fiorenzano}, A., {Harutyunyan}, A., {Maggio}, A., {Malavolta}, L., {Maldonado}, J., {Micela}, G., {Molinari}, E., {Pagano}, I., {Pedani}, M., {Piotto}, G., {Poretti}, E., {Scandariato}, G., {Sozzetti}, A., {Stoev}, H.}, The GAPS Programme at TNG. XXXII. The revealing non-detection of metastable He I in the atmosphere of the hot Jupiter WASP-80b, {\it A\&A}, {\bf 658}, 136\new{--149} (2022)

\item {dos Santos}, L.A., \new{{Ehrenreich}, D., {Bourrier}, V., {Allart}, R., {King}, G., {Lendl}, M., {Lovis}, C., {Margheim}, S., {Mel{\'e}ndez}, J., {Seidel}, J. V., {Sousa}, S. G.}, Search for helium in the upper atmosphere of the hot Jupiter WASP-127 b using Gemini/Phoenix, {\it A\&A}, {\bf 640}, 29\new{--33} (2020)

\item {Lam}, K.W.F., \new{{Faedi}, F., {Brown}, D.~J.~A., {Anderson}, D.~R., {Delrez}, L., {Gillon}, M., {H{\'e}brard}, G., {Lendl}, M., {Mancini}, L., {Southworth}, J., {Smalley}, B., {Triaud}, A.~H.~M., {Turner}, O.~D., {Hay}, K.~L., {Armstrong}, D.~J., {Barros}, S.~C.~C., {Bonomo}, A.~S., {Bouchy}, F., {Boumis}, P., {Collier Cameron}, A., {Doyle}, A.~P., {Hellier}, C., {Henning}, T., {Jehin}, E., {King}, G., {Kirk}, J., {Louden}, T., {Maxted}, P.~F.~L., {McCormac}, J.~J., {Osborn}, H.~P., {Palle}, E., {Pepe}, F., {Pollacco}, D., {Prieto-Arranz}, J., {Queloz}, D., {Rey}, J., {S{\'e}gransan}, D., {Udry}, S., {Walker}, S., {West}, R.~G., {Wheatley}, P.~J.}, From dense hot Jupiter to low-density Neptune: The discovery of WASP-127b, WASP-136b, and WASP-138b, {\it A\&A}, {\bf 599}, 3\new{--12} (2017)

\item {Seidel}, J.V., \new{{Lendl}, M., {Bourrier}, V., {Ehrenreich}, D., {Allart}, R., {Sousa}, S.~G., {Cegla}, H.~M., {Bonfils}, X., {Conod}, U., {Grandjean}, A., {Wyttenbach}, A., {Astudillo-Defru}, N., {Bayliss}, D., {Heng}, K., {Lavie}, B., {Lovis}, C., {Melo}, C., {Pepe}, F., {S{\'e}gransan}, D., {Udry}, S.}, Hot Exoplanet Atmospheres Resolved with Transit Spectroscopy (HEARTS). VI. Non-detection of sodium with HARPS on the bloated super-Neptune WASP-127b, {\it A\&A}, {\bf 643}, 45\new{--54} (2020)

\item {Casasayas-Barris}, N., \new{{Orell-Miquel}, J., {Stangret}, M., {Nortmann}, L., {Yan}, F., {Oshagh}, M., {Palle}, E., {Sanz-Forcada}, J., {L{\'o}pez-Puertas}, M., {Nagel}, E., {Luque}, R., {Morello}, G., {Snellen}, I.~A.~G., {Zechmeister}, M., {Quirrenbach}, A., {Caballero}, J.~A., {Ribas}, I., {Reiners}, A., {Amado}, P.~J., {Bergond}, G., {Czesla}, S., {Henning}, Th., {Khalafinejad}, S., {Molaverdikhani}, K., {Montes}, D., {Perger}, M., {S{\'a}nchez-L{\'o}pez}, A., {Sedaghati}, E.}, CARMENES detection of the Ca II infrared triplet and possible evidence of He I in the atmosphere of WASP-76b, {\it A\&A}, {\bf 654}, 163\new{--182} (2021)

\item {West}, R.G., \new{{Hellier}, C., {Almenara}, J. -M., {Anderson}, D.~R., {Barros}, S.~C.~C., {Bouchy}, F., {Brown}, D.~J.~A., {Collier Cameron}, A., {Deleuil}, M., {Delrez}, L., {Doyle}, A.~P., {Faedi}, F., {Fumel}, A., {Gillon}, M., {G{\'o}mez Maqueo Chew}, Y., {H{\'e}brard}, G., {Jehin}, E., {Lendl}, M., {Maxted}, P.~F.~L., {Pepe}, F., {Pollacco}, D., {Queloz}, D., {S{\'e}gransan}, D., {Smalley}, B., {Smith}, A.~M.~S., {Southworth}, J., {Triaud}, A.~H.~M.~J., {Udry}, S.}, Three irradiated and bloated hot Jupiters:. WASP-76b, WASP-82b, and WASP-90b, {\it A\&A}, {\bf 585}, 126\new{--132} (2016)

\item Hartman, J. D. \new{{Bakos}, G. {\'A}., {Sato}, B., {Torres}, G., {Noyes}, R.~W., {Latham}, D.~W., {Kov{\'a}cs}, G., {Fischer}, D.~A., {Howard}, A.~W., {Johnson}, J.~A., {Marcy}, G.~W., {Buchhave}, L.~A., {F{\"u}resz}, G., {Perumpilly}, G., {B{\'e}ky}, B., {Stefanik}, R.~P., {Sasselov}, D.~D., {Esquerdo}, G.~A., {Everett}, M., {Csubry}, Z., {L{\'a}z{\'a}r}, J., {Papp}, I., {S{\'a}ri}, P.}, HAT-P-18b and HAT-P-19b: Two Low-density Saturn-mass Planets Transiting Metal-rich K Stars, {\it ApJ}, {\bf 726}, 52\new{--67} (2011)

\item {Barrag{\'a}n}, O., \new{{Armstrong}, D.~J., {Gandolfi}, D., {Carleo}, I., {Vidotto}, A.~A., {Villarreal D'Angelo}, C., {Oklop{\v{c}}i{\'c}}, A., {Isaacson}, H., {Oddo}, D., {Collins}, K., {Fridlund}, M., {Sousa}, S.~G., {Persson}, C.~M., {Hellier}, C., {Howell}, S., {Howard}, A., {Redfield}, S., {Eisner}, N., {Georgieva}, I.~Y., {Dragomir}, D., {Bayliss}, D., {Nielsen}, L.~D., {Klein}, B., {Aigrain}, S., {Zhang}, M., {Teske}, J., {Twicken}, J.~D., {Jenkins}, J., {Esposito}, M., {Van Eylen}, V., {Rodler}, F., {Adibekyan}, V., {Alarcon}, J., {Anderson}, D.~R., {Akana Murphy}, J.~M., {Barrado}, D., {Barros}, S.~C.~C., {Benneke}, B., {Bouchy}, F., {Bryant}, E.~M., {Butler}, R.~P., {Burt}, J., {Cabrera}, J., {Casewell}, S., {Chaturvedi}, P., {Cloutier}, R., {Cochran}, W.~D., {Crane}, J., {Crossfield}, I., {Crouzet}, N., {Collins}, K.~I., {Dai}, F., {Deeg}, H.~J., {Deline}, A., {Demangeon}, O.~D.~S., {Dumusque}, X., {Figueira}, P., {Furlan}, E., {Gnilka}, C., {Goad}, M.~R., {Goffo}, E., {Guti{\'e}rrez-Canales}, F., {Hadjigeorghiou}, A., {Hartman}, Z., {Hatzes}, A.~P., {Harris}, M., {Henderson}, B., {Hirano}, T., {Hojjatpanah}, S., {Hoyer}, S., {Kab{\'a}th}, P., {Korth}, J., {Lillo-Box}, J., {Luque}, R., {Marmier}, M., {Mo{\v{c}}nik}, T., {Muresan}, A., {Murgas}, F., {Nagel}, E., {Osborne}, H.~L.~M., {Osborn}, A., {Osborn}, H.~P., {Palle}, E., {Raimbault}, M., {Ricker}, G.~R., {Rubenzahl}, R.~A., {Stockdale}, C., {Santos}, N.~C., {Scott}, N., {Schwarz}, R.~P., {Shectman}, S., {Raimbault}, M., {Seager}, S., {S{\'e}gransan}, D., {Serrano}, L.~M., {Skarka}, M., {Smith}, A.~M.~S., {{\v{S}}ubjak}, J., {Tan}, T.~G., {Udry}, S., {Watson}, C., {Wheatley}, P.~J., {West}, R., {Winn}, J.~N., {Wang}, S.~X., {Wolfgang}, A., {Ziegler}, C.}, The young HD 73583 (TOI-560) planetary system: two 10-M$_{{\ensuremath{\oplus}}}$ mini-Neptunes transiting a 500-Myr-old, bright, and active K dwarf, {MNRAS}, {\bf 514}, 1606\new{--1627} (2022)

\item Albrecht, S., \new{{Winn}, J. N., {Johnson}, J. A., {Howard}, A. W., {Marcy}, G. W., {Butler}, R. P., {Arriagada}, P., {Crane}, J. D., {Shectman}, S. A., {Thompson}, I. B., {Hirano}, T., {Bakos}, G., {Hartman}, J. D.}, Obliquities of Hot Jupiter Host Stars: Evidence for Tidal Interactions and Primordial Misalignments, {\it ApJ}, {\bf 757}, 18\new{--42} (2012)

\item Stassun, K. G. \new{{Collins}, K. A., {Gaudi}, B. S.}, Accurate Empirical Radii and Masses of Planets and Their Host Stars with Gaia Parallaxes, {\it AJ}, {\bf 153}, 136\new{--155} (2017)

\item Alonso-Floriano, F. J., \new{{Snellen}, I.~A.~G., {Czesla}, S., {Bauer}, F.~F., {Salz}, M., {Lamp{\'o}n}, M., {Lara}, L.~M., {Nagel}, E., {L{\'o}pez-Puertas}, M., {Nortmann}, L., {S{\'a}nchez-L{\'o}pez}, A., {Sanz-Forcada}, J., {Caballero}, J.~A., {Reiners}, A., {Ribas}, I., {Quirrenbach}, A., {Amado}, P.~J., {Aceituno}, J., {Anglada-Escud{\'e}}, G., {B{\'e}jar}, V.~J.~S., {Brinkm{\"o}ller}, M., {Hatzes}, A.~P., {Henning}, Th., {Kaminski}, A., {K{\"u}rster}, M., {Labarga}, F., {Montes}, D., {Pall{\'e}}, E., {Schmitt}, J.~H.~M.~M., {Zapatero Osorio}, M.~R.}, He I {\ensuremath{\lambda}} 10 830 {\r{A}} in the transmission spectrum of HD209458 b, {\it A\&A}, {\bf 629}, 110\new{--116} (2019)

\item Osborn, H. P., \new{{Bonfanti}, A., {Gandolfi}, D., {Hedges}, C., {Leleu}, A., {Fortier}, A., {Futyan}, D., {Gutermann}, P., {Maxted}, P.~F.~L., {Borsato}, L., {Collins}, K.~A., {Gomes da Silva}, J., {G{\'o}mez Maqueo Chew}, Y., {Hooton}, M.~J., {Lendl}, M., {Parviainen}, H., {Salmon}, S., {Schanche}, N., {Serrano}, L.~M., {Sousa}, S.~G., {Tuson}, A., {Ulmer-Moll}, S., {Van Grootel}, V., {Wells}, R.~D., {Wilson}, T.~G., {Alibert}, Y., {Alonso}, R., {Anglada}, G., {Asquier}, J., {Barrado y Navascues}, D., {Baumjohann}, W., {Beck}, T., {Benz}, W., {Biondi}, F., {Bonfils}, X., {Bouchy}, F., {Brandeker}, A., {Broeg}, C., {B{\'a}rczy}, T., {Barros}, S.~C.~C., {Cabrera}, J., {Charnoz}, S., {Collier Cameron}, A., {Csizmadia}, S., {Davies}, M.~B., {Deleuil}, M., {Delrez}, L., {Demory}, B. -O., {Ehrenreich}, D., {Erikson}, A., {Fossati}, L., {Fridlund}, M., {Gillon}, M., {G{\"o}mez-Munoz}, M.~A., {G{\"u}del}, M., {Heng}, K., {Hoyer}, S., {Isaak}, K.~G., {Kiss}, L., {Laskar}, J., {Lecavelier des Etangs}, A., {Lovis}, C., {Magrin}, D., {Malavolta}, L., {McCormac}, J., {Nascimbeni}, V., {Olofsson}, G., {Ottensamer}, R., {Pagano}, I., {Pall{\'e}}, E., {Peter}, G., {Piazza}, D., {Piotto}, G., {Pollacco}, D., {Queloz}, D., {Ragazzoni}, R., {Rando}, N., {Rauer}, H., {Reimers}, C., {Ribas}, I., {Demangeon}, O.~D.~S., {Smith}, A.~M.~S., {Sabin}, L., {Santos}, N., {Scandariato}, G., {Schroffenegger}, U., {Schwarz}, R.~P., {Shporer}, A., {Simon}, A.~E., {Steller}, M., {Szab{\'o}}, G.~M., {S{\'e}gransan}, D., {Thomas}, N., {Udry}, S., {Walter}, I., {Walton}, N.}, Uncovering the true periods of the young sub-Neptunes orbiting TOI-2076, {\it A\&A}, {\bf 664}, 156\new{--172} (2022)

\item Salz, M., \new{{Czesla}, S., {Schneider}, P.~C., {Nagel}, E., {Schmitt}, J.~H.~M.~M., {Nortmann}, L., {Alonso-Floriano}, F.~J., {L{\'o}pez-Puertas}, M., {Lamp{\'o}n}, M., {Bauer}, F.~F., {Snellen}, I.~A.~G., {Pall{\'e}}, E., {Caballero}, J.~A., {Yan}, F., {Chen}, G., {Sanz-Forcada}, J., {Amado}, P.~J., {Quirrenbach}, A., {Ribas}, I., {Reiners}, A., {B{\'e}jar}, V.~J.~S., {Casasayas-Barris}, N., {Cort{\'e}s-Contreras}, M., {Dreizler}, S., {Guenther}, E.~W., {Henning}, T., {Jeffers}, S.~V., {Kaminski}, A., {K{\"u}rster}, M., {Lafarga}, M., {Lara}, L.~M., {Molaverdikhani}, K., {Montes}, D., {Morales}, J.~C., {S{\'a}nchez-L{\'o}pez}, A., {Seifert}, W., {Zapatero Osorio}, M.~R., {Zechmeister}, M.}, Detection of He I {\ensuremath{\lambda}}10830 {\r{A}} absorption on HD 189733 b with CARMENES high-resolution transmission spectroscopy, {\it A\&A}, {\bf 620}, 97\new{--109} (2018)

\item Guilluy, G., \new{{Andretta}, V., {Borsa}, F., {Giacobbe}, P., {Sozzetti}, A., {Covino}, E., {Bourrier}, V., {Fossati}, L., {Bonomo}, A.~S., {Esposito}, M., {Giampapa}, M.~S., {Harutyunyan}, A., {Rainer}, M., {Brogi}, M., {Bruno}, G., {Claudi}, R., {Frustagli}, G., {Lanza}, A.~F., {Mancini}, L., {Pino}, L., {Poretti}, E., {Scandariato}, G., {Affer}, L., {Baffa}, C., {Baruffolo}, A., {Benatti}, S., {Biazzo}, K., {Bignamini}, A., {Boschin}, W., {Carleo}, I., {Cecconi}, M., {Cosentino}, R., {Damasso}, M., {Desidera}, S., {Falcini}, G., {Martinez Fiorenzano}, A.~F., {Ghedina}, A., {Gonz{\'a}lez-{\'A}lvarez}, E., {Guerra}, J., {Hernandez}, N., {Leto}, G., {Maggio}, A., {Malavolta}, L., {Maldonado}, J., {Micela}, G., {Molinari}, E., {Nascimbeni}, V., {Pagano}, I., {Pedani}, M., {Piotto}, G., {Reiners}, A.}, The GAPS programme at TNG. XXII. The GIARPS view of the extended helium atmosphere of HD 189733 b accounting for stellar activity, {\it A\&A}, {\bf 639}, 49 (2020)

\item Bakos, G. {\'A}. \new{{Torres}, G., {P{\'a}l}, A., {Hartman}, J., {Kov{\'a}cs}, G{\'e}za, {Noyes}, R.~W., {Latham}, D.~W., {Sasselov}, D.~D., {Sip{\H{o}}cz}, B., {Esquerdo}, G.~A., {Fischer}, D.~A., {Johnson}, J.~A., {Marcy}, G.~W., {Butler}, R.~P., {Isaacson}, H., {Howard}, A., {Vogt}, S., {Kov{\'a}cs}, G{\'a}bor, {Fernandez}, J., {Mo{\'o}r}, A., {Stefanik}, R.~P., {L{\'a}z{\'a}r}, J., {Papp}, I., {S{\'a}ri}, P.}, HAT-P-11b: A Super-Neptune Planet Transiting a Bright K Star in the Kepler Field, {\it ApJ}, {\bf 710}, 1724\new{--1745} (2010)

\item Yee, S. W., \new{{Petigura}, E. A., {Fulton}, B. J., {Knutson}, H. A., {Batygin}, K., {Bakos}, G., {Hartman}, J. D., {Hirsch}, L. A., {Howard}, A. W., {Isaacson}, H., {Kosiarek}, M. R., {Sinukoff}, E., {Weiss}, L. M.}, HAT-P-11: Discovery of a Second Planet and a Clue to Understanding Exoplanet Obliquities, {\it AJ}, {\bf 155}, 255\new{--267} (2018)

\item Awiphan, S. \new{{Kerins}, E., {Pichadee}, S., {Komonjinda}, S., {Dhillon}, V.~S., {Rujopakarn}, W., {Poshyachinda}, S., {Marsh}, T.~R., {Reichart}, D.~E., {Ivarsen}, K.~M., {Haislip}, J.~B.}, Transit timing variation and transmission spectroscopy analyses of the hot Neptune GJ3470b, {\it MNRAS}, {\bf 463}, 2574\new{--2582} (2016)

\item Palle, E., \new{{Nortmann}, L., {Casasayas-Barris}, N., {Lamp{\'o}n}, M., {L{\'o}pez-Puertas}, M., {Caballero}, J.~A., {Sanz-Forcada}, J., {Lara}, L.~M., {Nagel}, E., {Yan}, F., {Alonso-Floriano}, F.~J., {Amado}, P.~J., {Chen}, G., {Cifuentes}, C., {Cort{\'e}s-Contreras}, M., {Czesla}, S., {Molaverdikhani}, K., {Montes}, D., {Passegger}, V.~M., {Quirrenbach}, A., {Reiners}, A., {Ribas}, I., {S{\'a}nchez-L{\'o}pez}, A., {Schweitzer}, A., {Stangret}, M., {Zapatero Osorio}, M.~R., {Zechmeister}, M.}, A He I upper atmosphere around the warm Neptune GJ 3470 b, {\it A\&A}, {\bf 638}, 61\new{--68} (2020)

\item {Orell-Miquel}, J., \new{{Murgas}, F., {Pall{\'e}}, E., {Lamp{\'o}n}, M., {L{\'o}pez-Puertas}, M., {Sanz-Forcada}, J., {Nagel}, E., {Kaminski}, A., {Casasayas-Barris}, N., {Nortmann}, L., {Luque}, R., {Molaverdikhani}, K., {Sedaghati}, E., {Caballero}, J.~A., {Amado}, P.~J., {Bergond}, G., {Czesla}, S., {Hatzes}, A.~P., {Henning}, Th., {Khalafinejad}, S., {Montes}, D., {Morello}, G., {Quirrenbach}, A., {Reiners}, A., {Ribas}, I., {S{\'a}nchez-L{\'o}pez}, A., {Schweitzer}, A., {Stangret}, M., {Yan}, F., {Zapatero Osorio}, M.~R.}, A tentative detection of He I in the atmosphere of GJ 1214 b, {\it A\&A}, {\bf 659}, 55\new{--66} (2022)

\item {Petit dit de la Roche}, D.~J.~M., \new{{van den Ancker}, M.~E., {Miles-Paez}, P.~A.}, An Upper Limit on the Extended Helium Atmosphere of GJ 1214 b, {\it RNAAS}, {\bf 4}, 231 (2020)

\item {Cloutier}, R., \new{{Charbonneau}, D., {Deming}, D., {Bonfils}, X., {Astudillo-Defru}, N.}, A More Precise Mass for GJ 1214 b and the Frequency of Multiplanet Systems Around Mid-M Dwarfs, {\it AJ}, {\bf 162}, 174\new{--188} (2021)

\item Vissapragada, S., \new{{Knutson}, H. A., {Jovanovic}, N., {Harada}, C. K., {Oklop{\v{c}}i{\'c}}, A., {Eriksen}, J., {Mawet}, D., {Millar-Blanchaer}, M. A., {Tinyanont}, S., {Vasisht}, G.}, Constraints on Metastable Helium in the Atmospheres of WASP-69b and WASP-52b with Ultranarrowband Photometry, {\it AJ}, {\bf 159}, 278\new{--290} (2020)

\item Allart, R., \new{{Bourrier}, V., {Lovis}, C., {Ehrenreich}, D., {Aceituno}, J., {Guijarro}, A., {Pepe}, F., {Sing}, D.~K., {Spake}, J.~J., {Wyttenbach}, A.}, High-resolution confirmation of an extended helium atmosphere around WASP-107b, {\it A\&A}, {\bf 623}, 58\new{--63} (2019)

\item Kirk, J., \new{{Alam}, M. K., {L{\'o}pez-Morales}, M., {Zeng}, L.}, Confirmation of WASP-107b's Extended Helium Atmosphere with Keck II/NIRSPEC, {\it AJ}, {\bf 159}, 115\new{--123} (2020)

\item Mo{\v{c}}nik, T., \new{{Hellier}, C., {Anderson}, D.~R., {Clark}, B.~J.~M., {Southworth}, J.}, Starspots on WASP-107 and pulsations of WASP-118, {\it MNRAS}, {\bf 469}, 1622\new{--1629} (2017)

\item Piaulet, C. \new{{Benneke}, B., {Rubenzahl}, R. A., {Howard}, A. W., {Lee}, E. J., {Thorngren}, D., {Angus}, R., {Peterson}, M., {Schlieder}, J. E., {Werner}, M., {Kreidberg}, L., {Jaouni}, T., {Crossfield}, I. J.~M., {Ciardi}, D. R., {Petigura}, E. A., {Livingston}, J., {Dressing}, C. D., {Fulton}, B. J., {Beichman}, C., {Christiansen}, J. L., {Gorjian}, V., {Hardegree-Ullman}, K. K., {Krick}, J., {Sinukoff}, E.}, WASP-107b's Density Is Even Lower: A Case Study for the Physics of Planetary Gas Envelope Accretion and Orbital Migration, {\it AJ}, {\bf 161}, 70\new{--82} (2021)

\end{enumerate}

\section*{Acknowledgments}
These results are based on observations obtained with the Habitable-zone Planet Finder Spectrograph on the HET. The HET is a joint project of the University of Texas at Austin, the Pennsylvania State University, Ludwig-Maximilians-Universität München, and Georg-August Universität Gottingen. The HET is named in honor of its principal benefactors, William P. Hobby and Robert E. Eberly. The HET collaboration acknowledges the support and resources from the Texas Advanced Computing Center. We are grateful to the HET Resident Astronomers and Telescope Operators for their valuable assistance in gathering our HPF data.

\noindent {\bf Funding:}\\
NASA Exoplanets Research Program Grant number 80NSSC20K0257. (ZZ, CM, JL, MG) \\
Dutch Research Council NWO Veni Grant (AO) \\
National Science Foundation Grant 2108801 (WC) \\
NASA Hubble Fellowship grants HST-HF2-51522.001-A (ZZ), HST-HF2-51519.001-A (GS). \\
 The HPF team was also supported by the  NSF grants AST-1006676, AST-1126413, AST-1310875, AST-1310885, AST 2009889, AST 2009982, ATI 2009955, AAG 2108512, and the Heising-Simons Foundation via grant 2017-0494.

\noindent {\bf Author contributions:} \\
Data acquisition, proposal writing, and preparation of observations (JL, CM, MG, QT, JN, SM, DK, WC, BB, ME, GS, BT, AV) \\
Data reduction (GZ, MG, ZZ) \\
Observational data analysis (ZZ, CM, QT) \\
Hydrodynamical simulations (MM, AO, \new{JN}) \\
Paper writing and data visualization (ZZ, CM, MM, AO, MG, QT) \\
Paper review and editing (ZZ, CM, MG, MM, AO, QT, JL, JN, SM, DK, WC, BB, ME, GS, BT, AV, GZ)

\noindent {\bf Competing interests:} \\
Authors declare that they have no competing interests.

\sloppy
\noindent {\bf Data and materials availability:} 

\noindent \new{All data needed to evaluate the conclusions in the paper are present in the paper and/or the Supplementary Materials.} Availability of software used for data reduction of this work: the \texttt{Goldilocks} software is accessible at https://github.com/grzeimann/Goldilocks\_Documentation; the \texttt{muler} package is accessible at https://muler.readthedocs.io/en/latest/. Also, the HPF spectra, all measured spectral and physical properties, and our 3D hydrodynamic simulation results, as presented in this work are \new{all available on Zenodo: \url{https://doi.org/10.5281/zenodo.7767042}.}

%Here you should list the contents of your Supplementary Materials -- below is an example. 
%You should include a list of Supplementary figures, Tables, and any references that appear only in the SM. 
%Note that the reference numbering continues from the main text to the SM.
% In the example below, Refs. 4-10 were cited only in the SM.     
%\section*{Supplementary materials}
%Materials and Methods\\
%Figs. S1 to \new{S7} \\
%Tables S1 to S2  \\
%References 33--\new{115}

% For your review copy (i.e., the file you initially send in for
% evaluation), you can use the {figure} environment and the
% \includegraphics command to stream your figures into the text, placing
% all figures at the end.  For the final, revised manuscript for
% acceptance and production, however, PostScript or other graphics
% should not be streamed into your compliled file.  Instead, set
% captions as simple paragraphs (with a \noindent tag), setting them
% off from the rest of the text with a \clearpage as shown  below, and
% submit figures as separate files according to the Art Department's
% instructions.

\bigskip

\section*{Figures and Tables}

\clearpage

\begin{figure*}[t]
\begin{center}
\includegraphics[height=5.in]{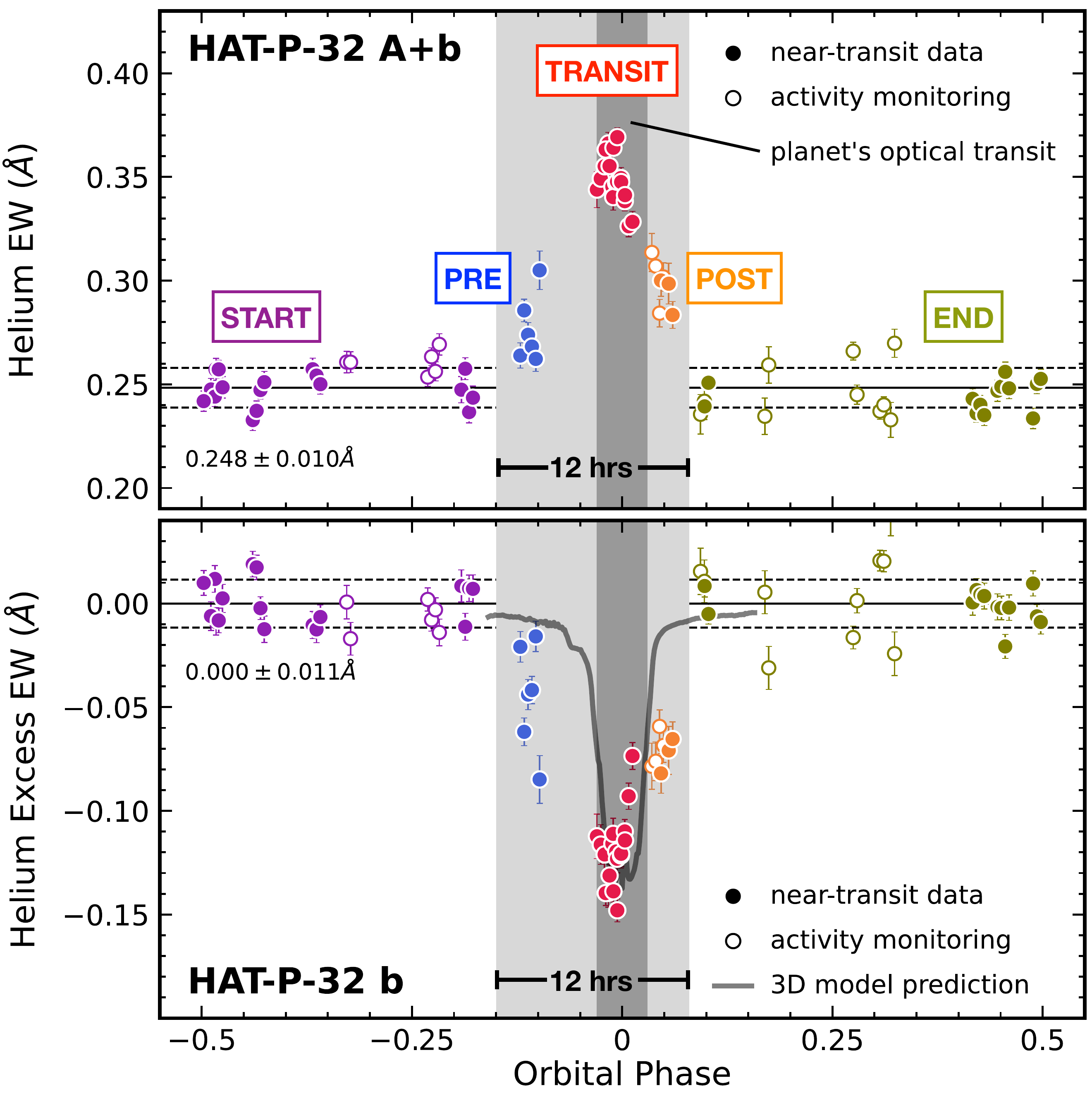}
\end{center}
\end{figure*}

\noindent {\bf Fig~1.} {\bf Equivalent widths (EWs) of the helium 10833~\AA\ triplet, measured from the observed spectra of HAT-P-32~A\new{+}b (top) and computed residual spectra of HAT-P-32~b (bottom), exhibit long-duration (12~hours), significant (14$\sigma$) excess near the planet's transits}. Solid circles represent data obtained within 2 nights of each optical transit and open circles represent the stellar variability monitoring data obtained from out-of-transit periods. We divided the data into five categories in terms of orbital phase, with boundaries highlighted by gray-shaded regions. \new{Our three-dimensional hydrodynamic simulation is shown as the gray solid line.} The time-series helium EWs are  asymmetric with respect to the planet's optical transit (dark gray), demonstrating the leading tail of the helium atmosphere escaping HAT-P-32~b.

\clearpage

\begin{figure*}[t]
\begin{center}
\includegraphics[height=4.8in]{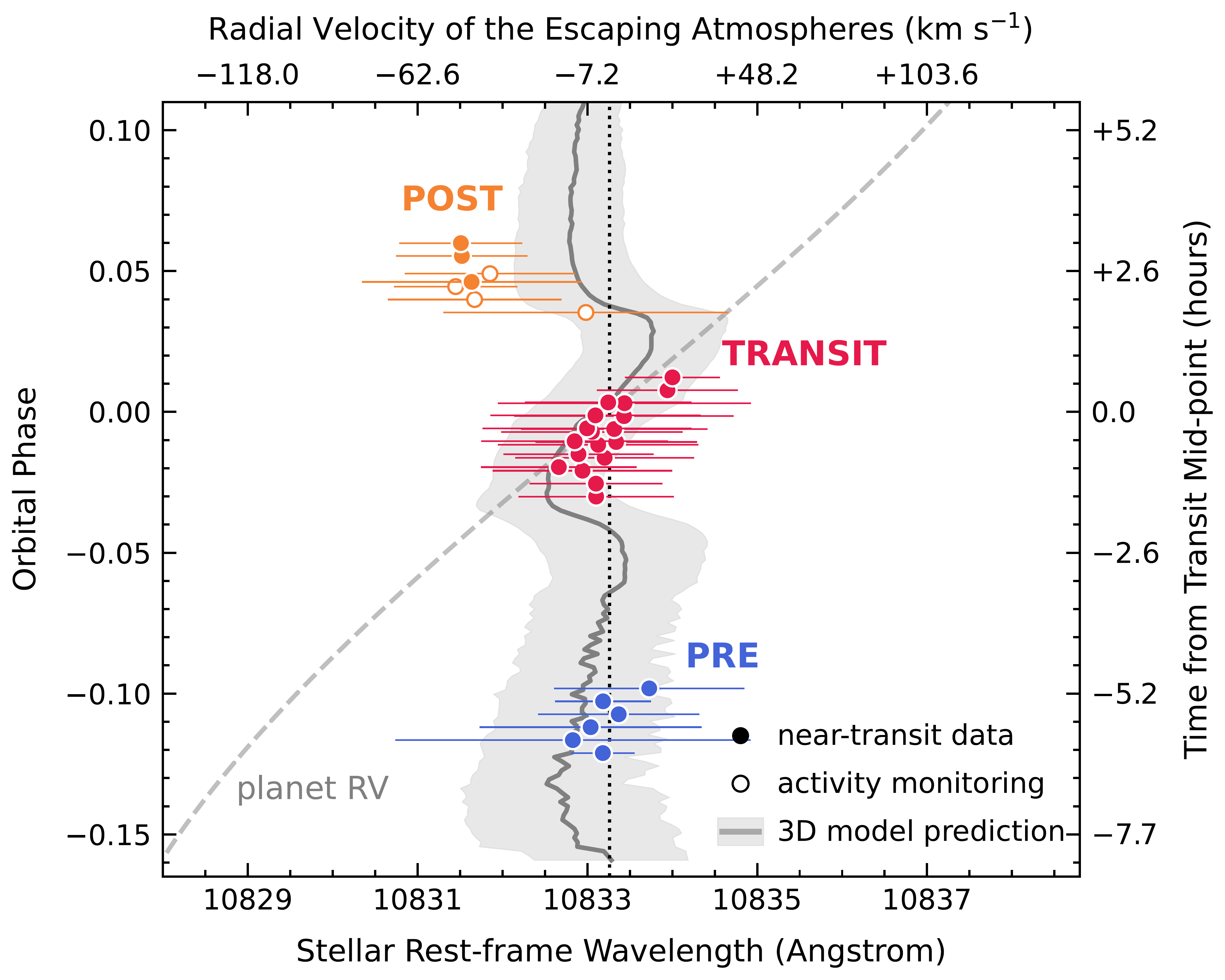}
\end{center}
\end{figure*}

\noindent {\bf Fig~2.} {\bf Measured central wavelengths and RVs of HAT-P-32~b's escaping helium atmosphere as a function of orbital phase in the stellar rest frame.} The observed helium excess features in \textsc{pre} (blue), \textsc{transit} (red), and \textsc{post} (orange) residual spectra were fitted by Gaussian profiles to determine the central wavelengths and the standard deviations (circles with horizontal bars), which were then converted to RVs of the escaping atmosphere based on the rest wavelength of the two strongest components of the helium triplet (vertical dotted line \new{at 10833.26~\AA}). Circle symbols have the same format as those shown in Fig. 1. The gray dashed line represents the planet's RV. \new{The central wavelength and standard deviation of helium excess from our simulations are shown as the gray solid line with shadow.}

\clearpage

\begin{figure*}[t]
\begin{center}
\includegraphics[height=5.6in]{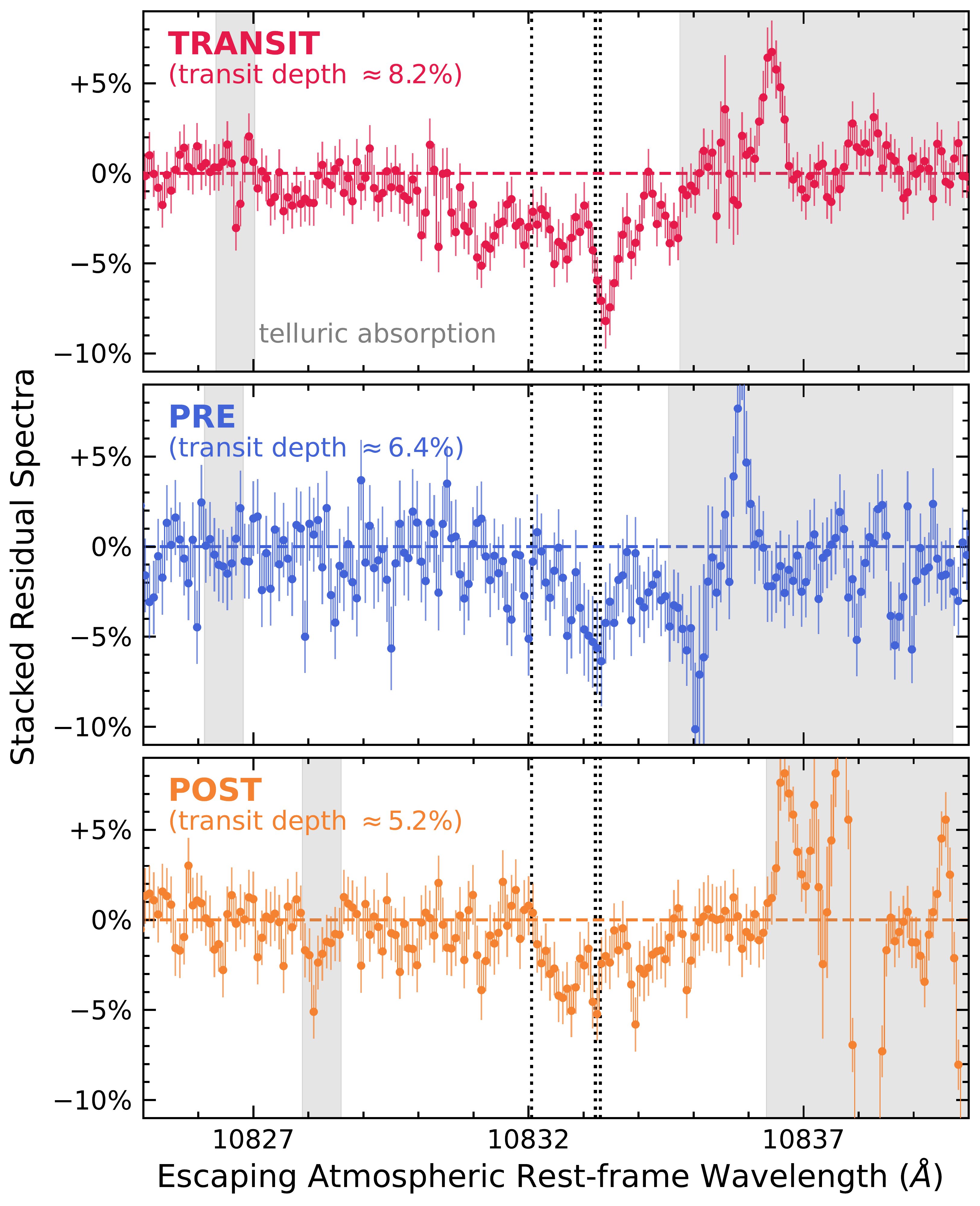}
\end{center}
\end{figure*}

\noindent {\bf Fig~3.} {\bf Stacked residual spectra of HAT-P-32~b.} \new{Before the stacking, each residual spectrum has been corrected by its RV of the escaping helium atmosphere, such that their helium excess features all line up with the rest wavelength of the strongest component of the helium triplet.} We detected strong excess helium absorption from \textsc{transit} (top), \textsc{pre} (middle), and \textsc{post} (bottom) residual spectra, spanning $1.5-4$~\AA\ in wavelength with a maximum depth of \new{$8.2\%$, $6.4\%$, and $5.2\%$}, respectively. The rest wavelength of the helium triplet is marked by vertical dotted lines. Telluric absorption features with transmission $<99.9\%$ are shown as gray shades.

\clearpage

\newpage

\begin{figure*}[t]
\begin{center}
\includegraphics[height=5.2in]{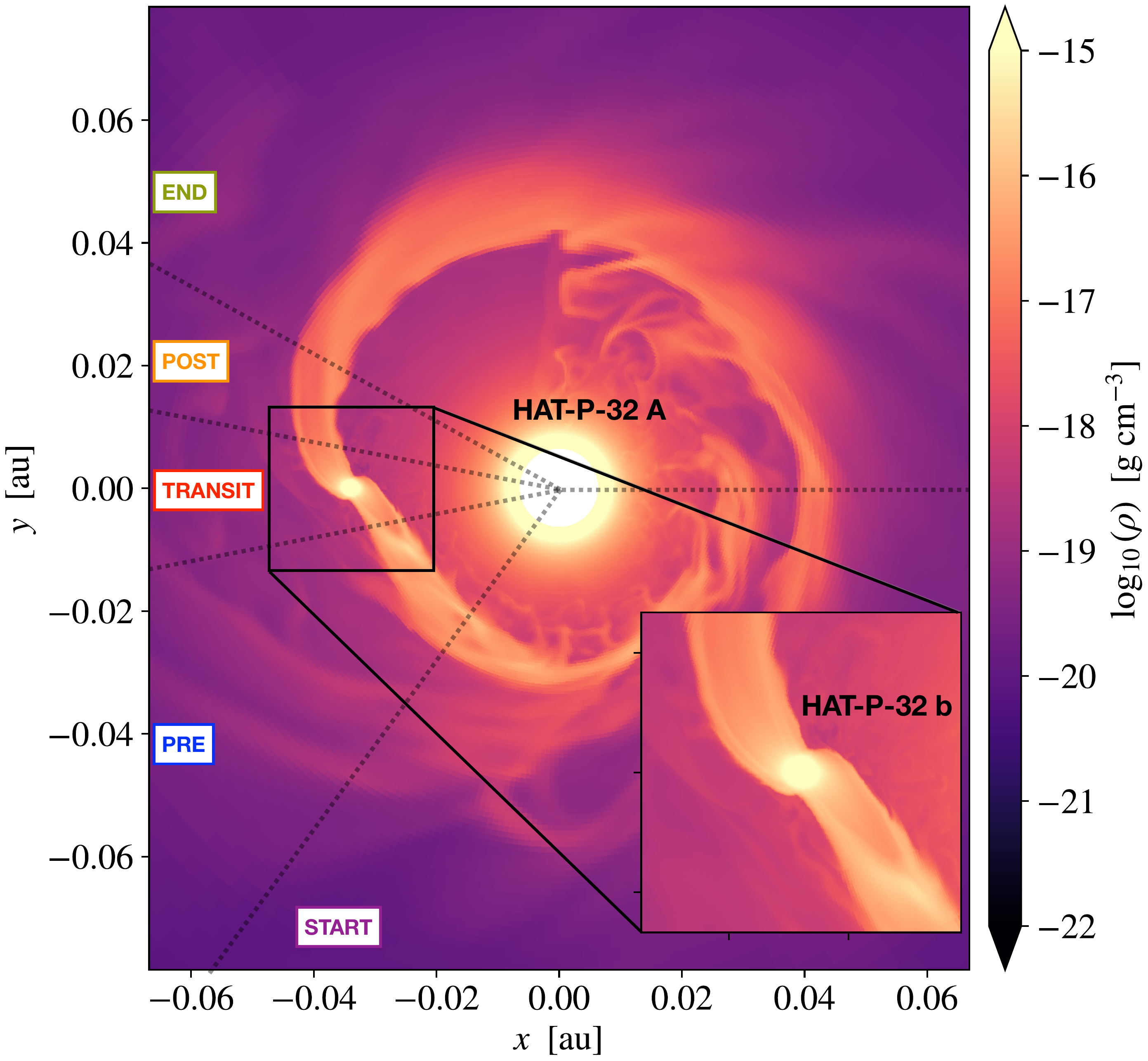}
\end{center}
\end{figure*}

\noindent {\bf Fig. 4.} {\bf Slice through the orbital plane of a simulated system approximating HAT-P-32~A\new{+}b. } The frame rotates with the planetary mean motion, so the position of observers rotates clockwise, with regions of \textsc{start}, \textsc{pre}, \textsc{transit}, \textsc{post}, and \textsc{end} shown in observations divided by black dotted lines. The logarithm of gas density is shown in the color scale. A low-density but relatively fast stellar wind expands from HAT-P-32~A at the coordinate origin and interacts with the outflow from HAT-P-32~b. The outflow from HAT-P-32~b is stretched into long, column-like tails leading and trailing the planet along the orbital path. These tidal tails are shaped by the advection of slow-moving planetary outflow in the star-planet gravitational field.

\newpage
\new{
\begin{figure*}[t]
\begin{center}
\includegraphics[height=5.in]{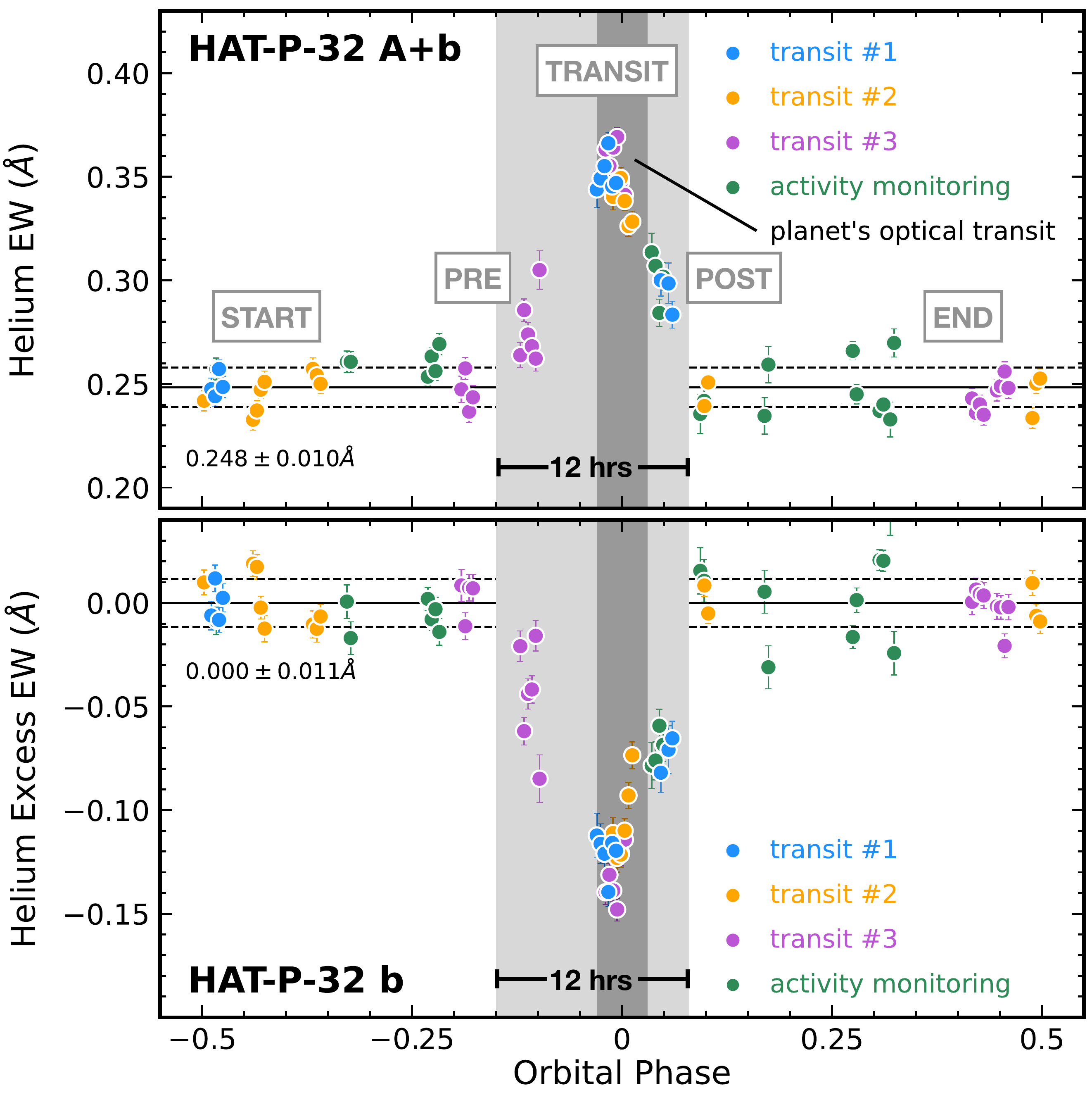}
\end{center}
\end{figure*}

\noindent {\bf Fig. 5.} {\bf The measured helium EWs of HAT-P-32~A+b and the helium excess EWs of HAT-P-32~b from different dates are consistent over a given range of the planet's orbital phase.} The data presented here are exactly the same as Fig. 1 but are color-coded by observation dates, including those observed during the first (blue), second (orange), and third (purple) transit event, as well as the long-term stellar activity monitoring (green). More observations slightly before the planet's optical transit are needed to examine whether the planet's helium excess is variable in the PRE subset.
}

\newpage

\begin{figure*}[t]
\begin{center}
\includegraphics[height=5.2in]{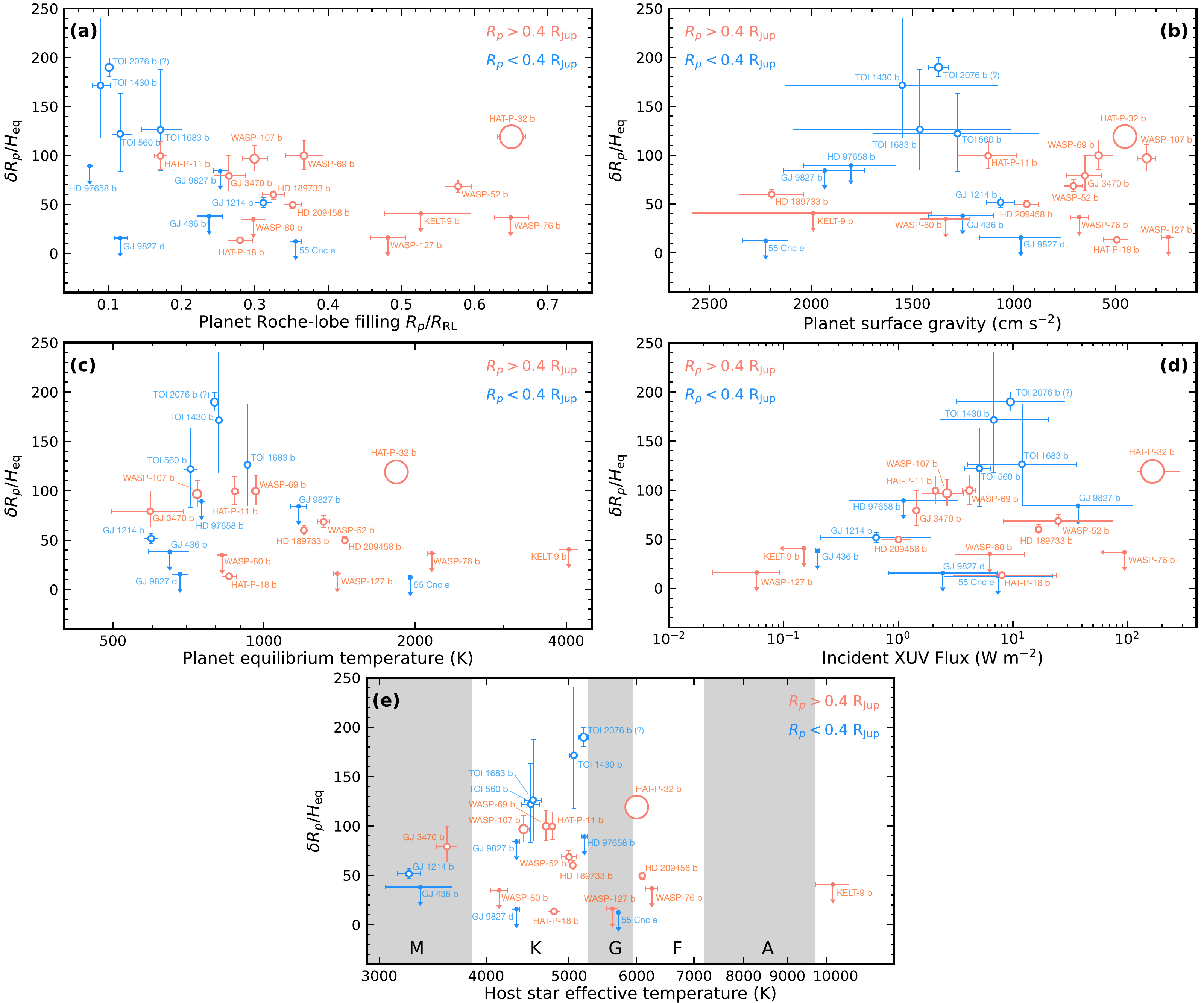}
\end{center}
\end{figure*}

\noindent {\bf Fig. \new{6}.} {\bf Census of planetary systems with detections and upper-limit constraints of excess helium absorption (\new{Table~S1}), divided into gas giants (orange) and sub-Neptunes (blue).} \new{The y-axes of all panels present the objects' equivalent heights of helium upper atmosphere in units of the scale height at equilibrium temperatures, $\delta{R_{p}}/H_{\rm eq}$. The x-axis in panel (a) is Roche-lobe filling $R_{p}/R_{\rm RL}$, which is the radius ratio between the planet and its Roche lobe. The x-axis in panels (b)--(e) represents these planets' surface gravities, equilibrium temperatures, incident XUV flux from their host stars, and their host stars' effective temperatures.} Symbol size for detections is proportional to the ratio between the transit duration of helium excess and that of the planets' optical transits. Most systems have such ratios as 1, while WASP-69~b ({\it 12}), WASP-107~b ({\it 28}), and HAT-P-32~b have larger ratios of 1.2, 1.4, and 3.8, respectively, meaning their helium upper atmospheres are extended. Also, TOI~2076~b exhibited excess helium absorption until 50~minutes after the planet's egress; we computed a ratio of 1.3 although the monitoring baseline did not extend the full optical transit and pre-ingress periods ({\it \new{36}}). \new{Furthermore, ({\it 37}) recently cautioned that the excess helium absorption signature of TOI~2076~b is likely due to its host star's variability.} In addition, TOI~1430~b exhibits excess helium absorption slightly before the planet's ingress ({\it 36}); longer-baseline monitoring would validate this feature so we simply assumed a ratio of 1.

\newpage

\begin{figure*}[t]
\begin{center}
\includegraphics[height=3.in]{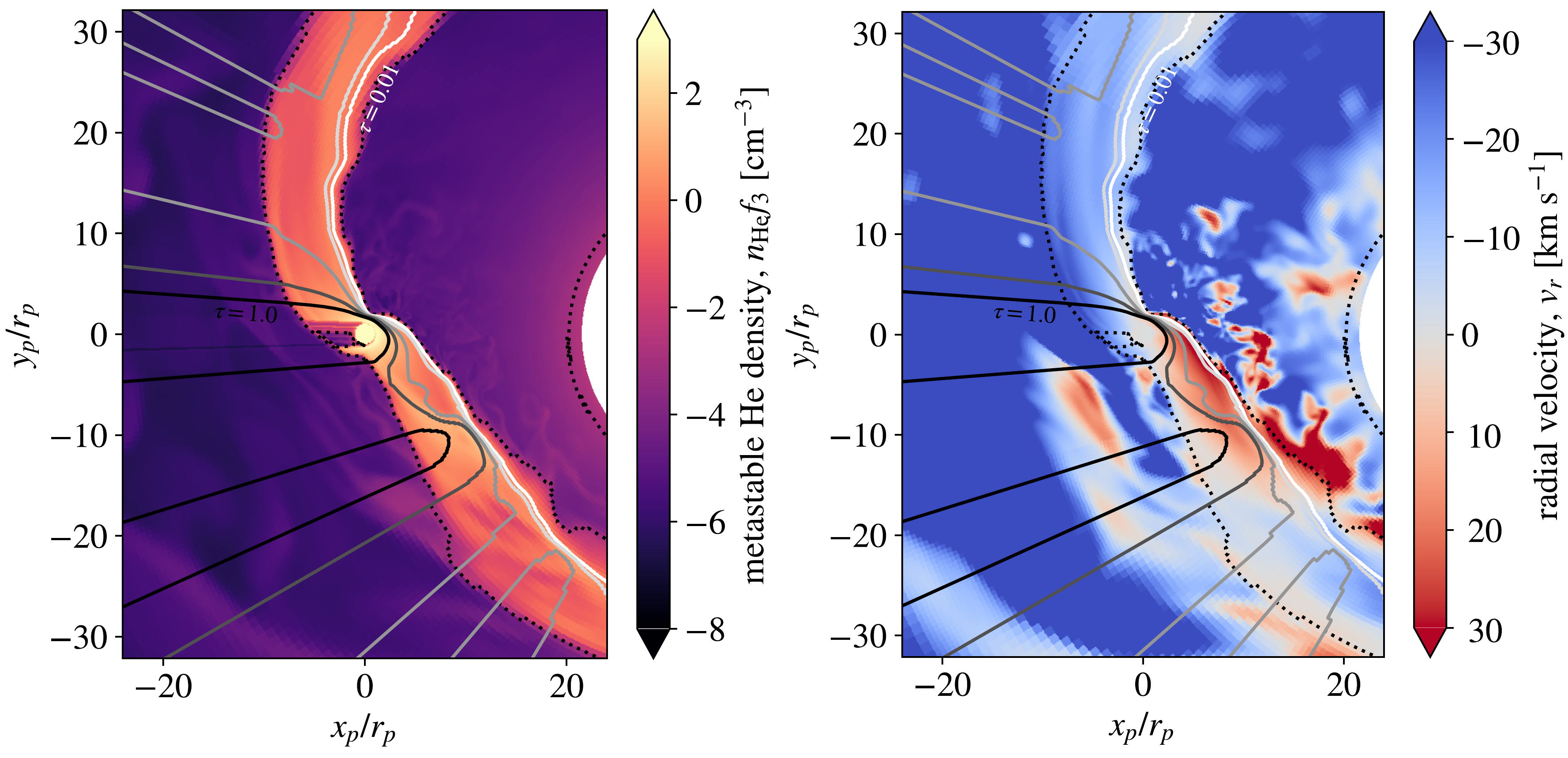}
\end{center}
\end{figure*}

\noindent \new{\bf Fig. 7.} {\bf Post-processed simulation snapshot (same as Figure 4.) showing number density of metastable helium (left) and radial velocity (right).  } Contours show the cumulative radial optical depths of 0.01 to 1 in half-dex steps (light to dark). The dotted contour shows the surface of $n_{\rm He}f_3=10^{-3}$~cm$^{-3}$.  The extended tidal tails of mass loss in this HAT-P-32~Ab analog system imply significant optical depth of the planetary mass loss even far from the planet itself, in significant contrast to a more-spherical pattern of mass loss. Further, these tidal tails orbit the host star with little radial velocity (or line-of-sight velocity for an observer) implying that their absorption signatures lie close to the stellar rest frame, which is in excellent agreement with observations (see Fig. 2).

\newpage

\begin{tabular}{llllll} 
\hline 
{\bf Date (UT)}  &  {\bf Type} &  {\bf $N_{\rm spec}$}  &  {\bf Mean S/N}  &  {\bf Airmass}  &  {\bf Orbital Phase}    \\ 
\hline 
\multicolumn{6}{c}{transit event 1}    \\ 
\hline 
2020 Aug 07  & \textsc{post} & 3  &  57  &  [1.31, 1.20]  &  [$+0.05$, $+0.06$]    \\ 
2020 Aug 08  & \textsc{start} & 4  &  91  &  [1.30, 1.19]  &  [$-0.49$, $-0.48$]    \\ 
2020 Aug 09  & \textsc{transit} & 6  &  68  &  [1.36, 1.16]  &  [$-0.03$, $-0.01$]    \\ 
\hline 
\multicolumn{6}{c}{transit event 2}    \\ 
\hline 
2020 Sept 18  & \textsc{start} & 3  &  89  &  [1.14, 1.20]  &  [$-0.37$, $-0.36$]    \\ 
2020 Sept 19  & \textsc{transit} & 6  &  84  &  [1.35, 1.16]  &  [$-0.01$, $+0.01$]    \\ 
2020 Sept 19  & \textsc{end} & 2  &  112  &  [1.16, 1.18]  &  $+0.10$    \\ 
2020 Sept 20  & \textsc{start} & 4  &  93  &  [1.15, 1.24]  &  [$-0.44$, $-0.43$]    \\ 
2020 Sept 22  & \textsc{end}+\textsc{start} & 4  &  95  &  [1.15, 1.24]  &  [$+0.49$, $+0.51$]    \\ 
\hline 
\multicolumn{6}{c}{transit event 3}    \\ 
\hline 
2020 Oct 5  & \textsc{end} & 4  &  96  &  [1.29, 1.18]  &  [$+0.42$, $+0.43$]    \\ 
2020 Oct 6  & \textsc{pre} & 6  &  74  &  [1.31, 1.14]  &  [$-0.12$, $-0.10$]    \\
2020 Oct 6  & \textsc{transit} & 6  &  93  &  [1.14, 1.31]  &  [$-0.02$, $0.00$]    \\ 
2020 Oct 7  & \textsc{end} & 4  &  96  &  [1.15, 1.24]  &  [$+0.45$, $+0.46$]    \\ 
2020 Oct 8  & \textsc{start} & 4  &  83  &  [1.29, 1.18]  &  [$-0.19$, $-0.18$]    \\ 
\hline 
\multicolumn{6}{c}{stellar activity monitoring}    \\ 
\hline 
2020 Aug 01  & \textsc{end} & 2  &  99  &  [1.21, 1.18]  &  $+0.28$    \\ 
2020 Sept 04  & \textsc{post} & 4  &  59  &  [1.31, 1.19]  &  [$+0.04$, $+0.05$]    \\ 
2020 Sept 05  & \textsc{start} & 1  &  84  &  1.17  &  $-0.48$    \\ 
2020 Oct 11  & \textsc{end} & 2  &  116  &  [1.18, 1.21]  &  $+0.31$    \\ 
2020 Oct 12  & \textsc{start} & 4  &  100  &  [1.17, 1.27]  &  [$-0.23$, $-0.22$]    \\ 
2020 Dec 02  & \textsc{end} & 2  &  66  &  [1.30, 1.26]  &  $+0.32$    \\ 
2020 Dec 09  & \textsc{start} & 2  &  90  &  [1.17, 1.20]  &  [$-0.33$, $-0.32$]    \\ 
2020 Dec 23  & \textsc{end} & 2  &  71  &  [1.20, 1.23]  &  $+0.17$    \\ 
2020 Dec 25  & \textsc{end} & 2  &  74  &  [1.17, 1.20]  &  [$+0.09$, $+0.10$]    \\ 
\hline 
\end{tabular} 

\bigskip
\bigskip

\noindent {\bf Table. 1.} {\bf The HET/HPF observing log of HAT-P-32~Ab.} For each date, we list the number of the observed spectra ($N_{\rm spec}$) and their mean S/N (per pixel) near 10833~\AA, as well as the range of airmass and orbital phase covered by these data. \new{Each spectrum was acquired with an exposure time of 820~sec. The HPF spectral resolution element contains a median of 2.8 pixels.} We also divide our spectra into five subsets based on their orbital phase as indicated in the ``Type'' column, including \textsc{start} (with the orbital phase in $[-0.5, -0.15]$), \textsc{pre} ($[-0.15, -0.03]$), \textsc{transit} ($[-0.03, +0.03]$), \textsc{post} ($[+0.03, +0.08]$), and \textsc{end} ($[+0.08, +0.5]$).

\newpage

\begin{center}
{\Large Supplementary Materials for

\bigskip
\bigskip

Giant Tidal Tails of Helium Escaping \\the Hot Jupiter HAT-P-32~b
}

\bigskip

\author
{Zhoujian Zhang,$^{1,2,\ast}$ Caroline V. Morley,$^{2}$\\ Michael Gully-Santiago,$^{2}$ Morgan MacLeod,$^{3}$ Antonija Oklop{\v{c}}i{\'c},$^{4}$\\ Jessica Luna,$^{2}$ Quang H. Tran,$^{2}$ Joe P. Ninan,$^{5}$\\ Suvrath Mahadevan,$^{6,7,8}$ Daniel M. Krolikowski,$^{2,9}$ William D. Cochran,$^{10}$\\ Brendan P. Bowler,$^{2}$ Michael Endl,$^{11}$ Gudmundur Stef{\'a}nsson,$^{12}$\\ Benjamin M. Tofflemire,$^{2}$ Andrew Vanderburg,$^{13}$ Gregory R. Zeimann$^{14}$ \\

\normalsize{$^{1}$Department of Astronomy \& Astrophysics, University of California, Santa Cruz, CA 95064, USA}\\
\normalsize{$^{2}$Department of Astronomy, The University of Texas at Austin, Austin, TX 78712, USA}\\
\normalsize{$^{3}$Center for Astrophysics, Harvard \& Smithsonian, Cambridge, MA 02138, USA}\\
\normalsize{$^{4}$Anton Pannekoek Institute for Astronomy, University of Amsterdam, The Netherlands}\\
\normalsize{$^{5}$Department of Astronomy \& Astrophysics, Tata Institute of Fundamental Research, India}\\
\normalsize{$^{6}$Department of Astronomy \& Astrophysics, The Pennsylvania State University, USA}\\
\normalsize{$^{7}$Center for Exoplanets and Habitable Worlds, USA}\\
\normalsize{$^{8}$ETH Zurich, Institute for Particle Physics \& Astrophysics, Zurich, Switzerland}\\
\normalsize{$^{9}$\new{Steward Observatory, The University of Arizona, 933 N. Cherry Ave, Tucson, AZ 85721, USA}}\\
\normalsize{$^{10}$Center for Planetary Systems Habitability and McDonald Observatory, UT Austin, USA}\\
\normalsize{$^{11}$McDonald Observatory and the Department of Astronomy, UT Austin, USA}\\
\normalsize{$^{12}$Princeton University, Department of Astrophysical Sciences, USA}\\
\normalsize{$^{13}$Department of Physics and Kavli Institute for Astrophysics and Space Research, MIT, USA}\\
\normalsize{$^{14}$Hobby-Eberly Telescope, The University of Texas at Austin, Austin, Austin, TX, 78712, USA}\\

\normalsize{$^\ast$To whom correspondence should be addressed; E-mail:  zhangdirac [at] gmail [dot] com} 
}
\end{center}

\noindent {\bf This PDF file includes:}

Figs. S1 to \new{S4} 

Table \new{S1}  

References \new{62--112}

Data \new{S1}

\newpage

\begin{figure*}[t]
\begin{center}
\includegraphics[height=5.6in]{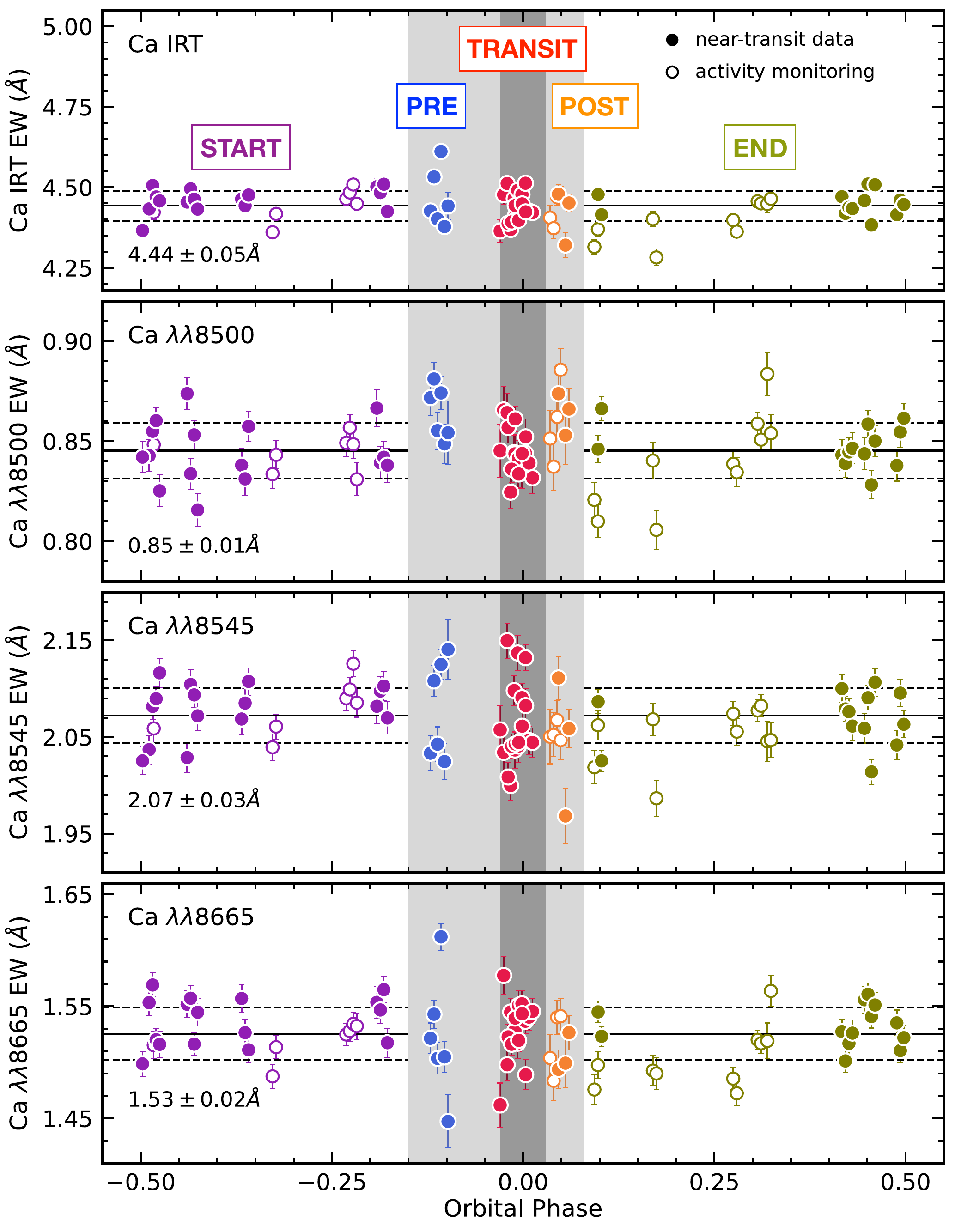}
\end{center}
\end{figure*}

\noindent {\bf Fig. S1.} {\bf Measured equivalent widths (EWs) of the calcium infrared triplet are not correlated with the planet's orbital phase, suggesting the observed excess helium absorption originates from the planet's exospheres rather than stellar activity indicators.} The top panel presents the total EWs of the Ca triplet, computed as the sum of individual components' EWs as shown in the remaining three panels. The format is the same as Fig. 1, with the weighted mean and root-mean-square of EWs during the \textsc{start} and \textsc{end} phases labeled at the bottom left of each panel.

\newpage

\begin{figure*}[t]
\begin{center}
\includegraphics[height=5in]{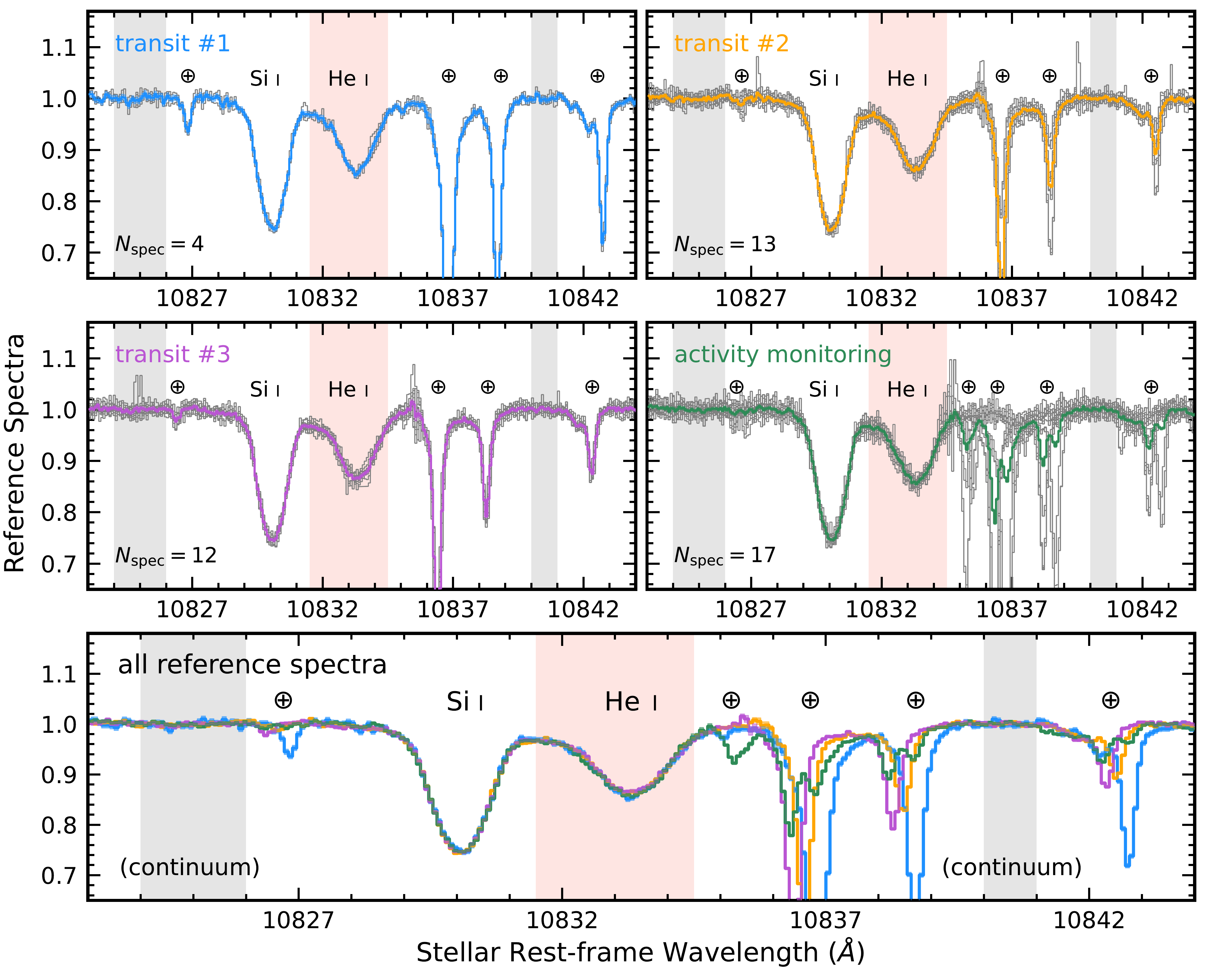}
\end{center}
\end{figure*}

\noindent {\bf Fig. S2.} {\bf Reference spectra for the first (top left; blue), second (top right; orange), and third (middle left; purple) transit event, as well as the long-term stellar activity monitoring (middle right; green).} In each panel, we label the number of spectra used for computing the reference spectra ($N_{\rm spec}$) and overlay individual spectra as gray lines. All four reference spectra have consistent shapes and fluxes near the helium and silicon features (bottom panel), with different strengths of telluric absorption (labeled by ``$\oplus$'') that are widely separated from the helium triplet at 10833~\AA. \new{In each panel, the red and gray shades show the wavelength ranges used for computing the line flux and the pseudo-continuum of the helium EW, respectively.}

\clearpage

\newpage

\begin{figure*}[t]
\begin{center}
\includegraphics[height=3.3in]{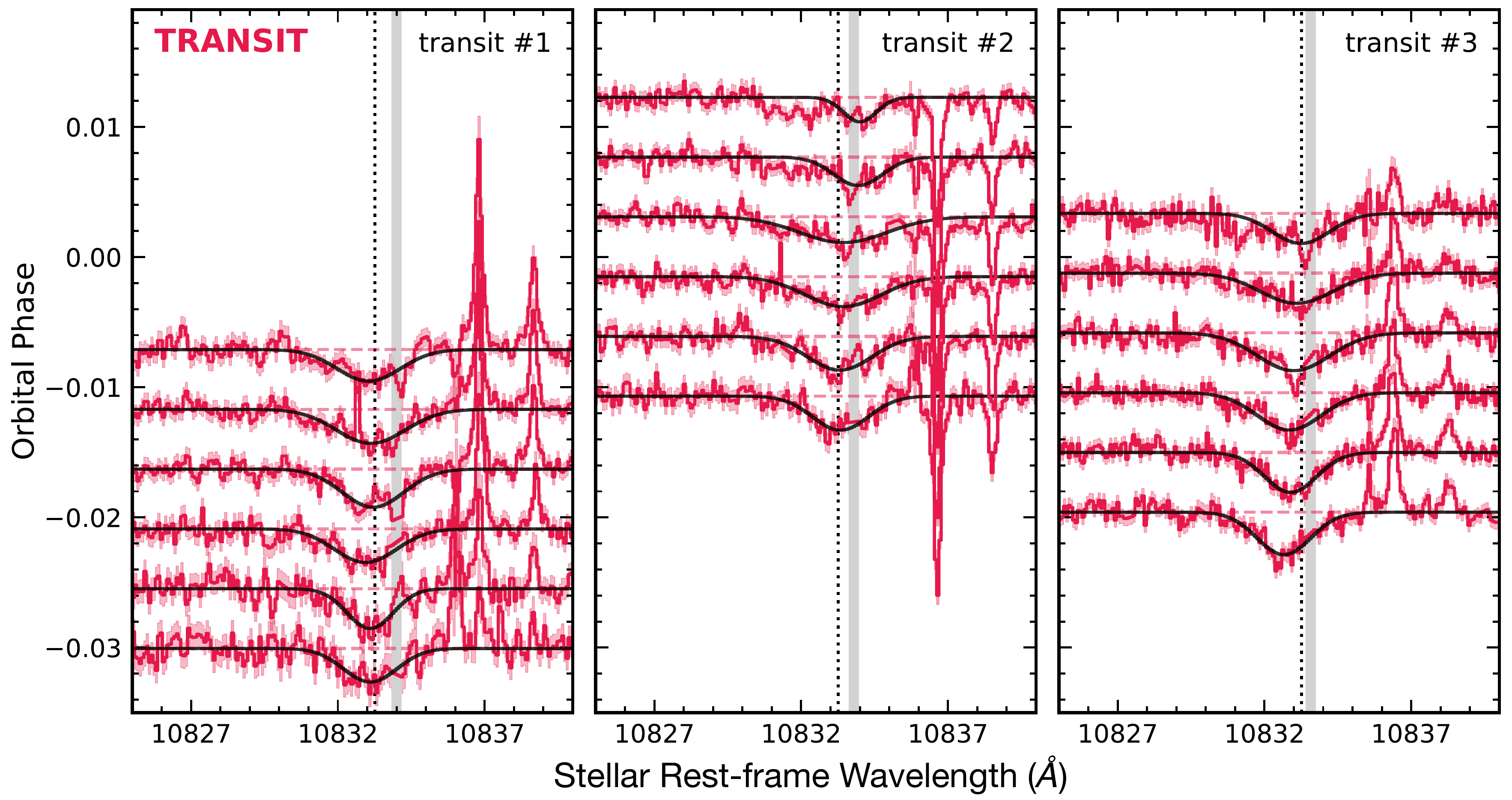}
\end{center}
\end{figure*}

\noindent {\bf Fig. S3.} {\bf Residual spectra (solid lines; scaled by a factor of 0.05) and $1\sigma$ uncertainties (shades) in the \textsc{transit} subset of all three transit events ordered by the orbital phase, showing significant excess helium absorption features.} We marked the rest wavelength of the two strongest and blended components of the helium triplet at 10833.26~\AA\ (dotted vertical line). \new{Vertical gray shades highlight the wavelength ranges of the OH skyline doublet, over which we masked and approximated the spectral fluxes based on linear interpolation.} A Gaussian fit to the Helium excess in each residual spectrum is overlaid (black). Sharp spikes and absorption features near 10837~\AA\ are due to the telluric absorption features that are widely separated from the helium signal.

\newpage

\begin{figure*}[t]
\begin{center}
\includegraphics[height=3.in]{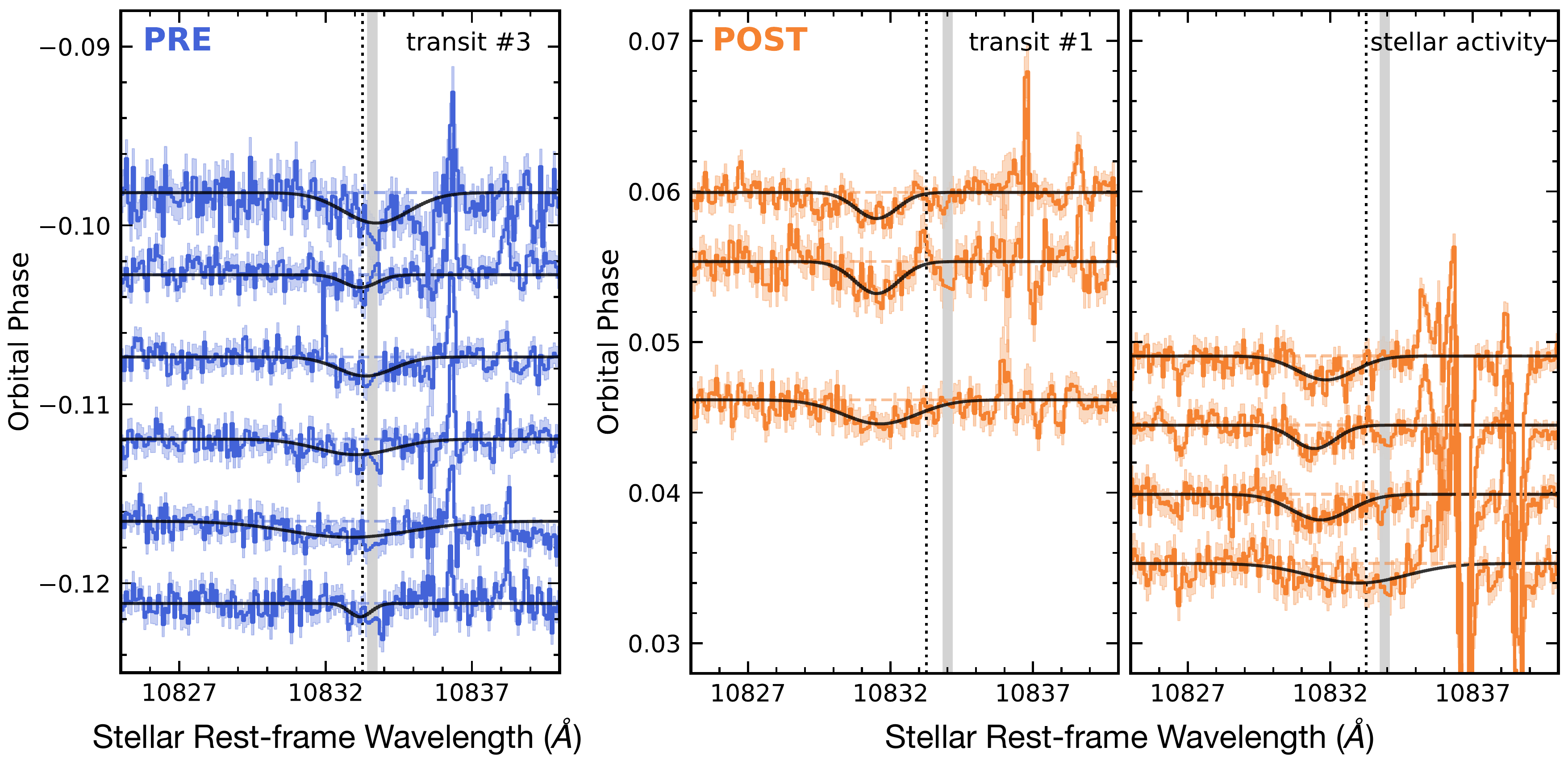}
\end{center}
\end{figure*}

\noindent {\bf Fig. S4.} {\bf Residual spectra (solid lines; scaled by a factor of 0.05) and $1\sigma$ uncertainties (shades) in the \textsc{pre} (left; orbital phase in $[-0.15, -0.03]$) and \textsc{post} (right; orbital phase in $[+0.03, +0.08]$) subsets ordered by the orbital phase.} The \textsc{pre} residual spectra are only from the third transit event, and the \textsc{post} residual spectra are from both the first transit event and our long-term stellar variability monitoring. The format is the same as Fig. S3.

\newpage

{\scriptsize 
\begin{tabular}{lccccccl} 
\hline 
{\bf System}  &  {\bf $\delta R_{p}/H_{\rm eq}$}  &  {\bf $R_{p}/R_{\rm RL}$}  &  {\bf Planet}  &  {\bf Planet}  &  {\bf Incident}  &  {\bf Host Star}  &  {\bf Reference}    \\  
{\bf }  &  {\bf }  &  {\bf }  &  {\bf surface gravity}  &  {\bf $T_{\rm eq}$}  &  {\bf XUV Flux}  &  {\bf $T_{\rm eff}$}  &  {\bf }    \\ 
{}  &  {}  &  {}  &  {(g cm$^{-2}$)}  &  {(K)}  &  {(W m$^{-2}$)}  &  {\bf (K)}  &  {\bf}   \\ 
\hline 
K2-136 c  &  --  &  --  &  --  &  $425 \pm 21$  &  $0.960 \pm 0.050$  &  $4499 \pm 50$  &  ({\it \new{62--65}}) \\ 
K2-100 b  &  --  &  $0.294 \pm 0.035$  &  $1419 \pm 418$  &  $1841 \pm 41$  &  $35.700 \pm 6.400$  &  $5945 \pm 110$  &  ({\it \new{66--67}})  \\ 
V1298 Tau c  &  --  &  --  &  --  &  $968 \pm 31$  &  $329.100 \pm 11.550$  &  $4970 \pm 120$  &   ({\it \new{68--70}}) \\ 
TRAPPIST-1 b  &  --  &  $0.023 \pm 0.001$  &  $1080 \pm 55$  &  $400 \pm 7$  &  $3.000 \pm 0.400$  &  $2559 \pm 50$  &  ({\it \new{71--73}}) \\ 
TRAPPIST-1 e  &  --  &  $0.093 \pm 0.009$  &  $801 \pm 33$  &  $251 \pm 4$  &  $0.400 \pm 0.070$  &  $2559 \pm 50$  &  ({\it \new{71--74}})  \\ 
TRAPPIST-1 f  &  --  &  $0.070 \pm 0.002$  &  $932 \pm 35$  &  $219 \pm 4$  &  $0.270 \pm 0.040$  &  $2559 \pm 50$  &  ({\it \new{71--74}})  \\ 
WASP-177 b  &  --  &  $0.483 \pm 0.155$  &  $494 \pm 298$  &  $1142 \pm 32$  &  $3.500 \pm 4.650$  &  $5017 \pm 70$  &  ({\it \new{75--76}})  \\ 
GJ 9827 d  &  $< 15$  &  $0.116 \pm 0.009$  &  $966 \pm 200$  &  $680 \pm 25$  &  $2.450 \pm 3.265$  &  $4340 \pm 46$  &  ({\it \new{33, 77--78}}) \\ 
V1298 Tau b  &  --  &  --  &  --  &  $677 \pm 22$  &  $78.600 \pm 2.750$  &  $4970 \pm 120$  &   ({\it \new{68--70}})  \\ 
HD 97658 b  &  $< 89$  &  $0.075 \pm 0.004$  &  $1803 \pm 227$  &  $751 \pm 12$  &  $1.110 \pm 1.480$  &  $5212 \pm 43$  &  ({\it \new{77,79}})  \\ 
55 Cnc e  &  $< 12$  &  $0.356 \pm 0.008$  &  $2223 \pm 111$  &  $1958 \pm 15$  &  $7.400 \pm 9.850$  &  $5712 \pm 18$  &  ({\it \new{80--82}})  \\ 
KELT-9 b  &  $< 40$  &  $0.527 \pm 0.059$  &  $1989 \pm 589$  &  $4050 \pm 180$  &  $<0.150$  &  $10170 \pm 450$  &  ({\it 12, \new{83}}) \\ 
GJ 9827 b  &  $< 84$  &  $0.253 \pm 0.010$  &  $1932 \pm 200$  &  $1172 \pm 43$  &  $37.000 \pm 49.500$  &  $4340 \pm 46$  &  ({\it \new{33, 77--78}})   \\ 
GJ 436 b  &  $< 38$  &  $0.238 \pm 0.018$  &  $1252 \pm 160$  &  $649 \pm 60$  &  $0.197 \pm 0.007$  &  $3350 \pm 300$  &  ({\it 12, \new{84--85}})  \\ 
WASP-80 b  &  $< 34$  &  $0.298 \pm 0.017$  &  $1336 \pm 121$  &  $825 \pm 19$  &  $6.281 \pm 4.711$  &  $4143 \pm 93$  &  ({\it \new{86--87}})  \\ 
WASP-127 b  &  $< 16$  &  $0.482 \pm 0.024$  &  $238 \pm 29$  &  $1400 \pm 24$  &  $0.058 \pm 0.034$  &  $5620 \pm 85$  &  ({\it \new{88--90}})  \\ 
WASP-76 b  &  $< 36$  &  $0.649 \pm 0.024$  &  $678 \pm 42$  &  $2160 \pm 40$  &  $<94.000$  &  $6250 \pm 100$  &  ({\it \new{91--92}})  \\ 
HAT-P-18 b  &  $13 \pm 2 $  &  $0.280 \pm 0.017$  &  $493 \pm 60$  &  $852 \pm 28$  &  $8.000 \pm 10.500$  &  $4803 \pm 80$  &  ({\it \new{39, 84, 93}})  \\ 
TOI 1430 b  &  $171 \pm 61 $  &  $0.089 \pm 0.013$  &  $1550 \pm 525$  &  $813$  &  $6.800 \pm 9.050$  &  $5067 \pm 60$  &  ({\it \new{36}}) \\ 
V1298 Tau d  &  --  &  --  &  --  &  $845 \pm 27$  &  $191.000 \pm 6.650$  &  $4970 \pm 120$  &   ({\it \new{68--70}})  \\ 
TOI 560 b  &  $121 \pm 40 $  &  $0.116 \pm 0.013$  &  $1278 \pm 408$  &  $714 \pm 21$  &  $5.100 \pm 1.300$  &  $4511 \pm 110$  &  ({\it \new{34, 94}})  \\ 
TOI 1683 b  &  $126 \pm 51 $  &  $0.171 \pm 0.028$  &  $1463 \pm 535$  &  $927$  &  $12.000 \pm 16.000$  &  $4539 \pm 100$  &  ({\it \new{36}})  \\ 
HD 209458 b  &  $49 \pm 3 $  &  $0.352 \pm 0.012$  &  $936 \pm 57$  &  $1449 \pm 12$  &  $1.004 \pm 0.284$  &  $6091 \pm 10$  &  ({\it 12, \new{84}, \new{95--97}})  \\ 
TOI 2076 b  &  $189 \pm 9 $  &  $0.101 \pm 0.003$  &  $1371 \pm 48$  &  $797 \pm 12$  &  $9.500 \pm 12.650$  &  $5200 \pm 70$  &  ({\it \new{36, 98}})  \\ 
HD 189733 b  &  $59 \pm 4 $  &  $0.325 \pm 0.015$  &  $2194 \pm 159$  &  $1201 \pm 12$  &  $16.750 \pm 0.028$  &  $5052 \pm 16$  &  ({\it 12, \new{84, 96, 99--100}}  \\ 
HAT-P-11 b  &  $99 \pm 14 $  &  $0.171 \pm 0.009$  &  $1127 \pm 145$  &  $875 \pm 15$  &  $2.109 \pm 0.124$  &  $4780 \pm 50$  &  ({\it 11, \new{101--102}})  \\ 
GJ 3470 b  &  $79 \pm 18 $  &  $0.265 \pm 0.021$  &  $650 \pm 86$  &  $593 \pm 96$  &  $1.435 \pm 0.008$  &  $3600 \pm 100$  &  ({\it \new{61, 103--104}})  \\ 
GJ 1214 b  &  $51 \pm 5 $  &  $0.312 \pm 0.011$  &  $1065 \pm 70$  &  $596 \pm 19$  &  $0.640 \pm 0.855$  &  $3250 \pm 100$  &   ({\it \new{77, 103--104}}) \\ 
WASP-52 b  &  $68 \pm 6 $  &  $0.578 \pm 0.018$  &  $707 \pm 45$  &  $1315 \pm 35$  &  $24.800 \pm 33.150$  &  $5000 \pm 100$  &  ({\it \new{35, 76, 107}})  \\ 
WASP-69 b  &  $99 \pm 14 $  &  $0.367 \pm 0.026$  &  $583 \pm 73$  &  $963 \pm 18$  &  $4.170 \pm 0.566$  &  $4700 \pm 50$  &  ({\it 12, \new{96, 108}})  \\ 
WASP-107 b  &  $96 \pm 13 $  &  $0.299 \pm 0.017$  &  $345 \pm 43$  &  $736 \pm 17$  &  $2.664 \pm 1.050$  &  $4425 \pm 70$  &  ({\it 10, 12, 28, \new{109--112}})  \\ 
HAT-P-32 b  &  \new{$119 \pm 8 $}  &  $0.650 \pm 0.019$  &  $453 \pm 27$  &  $1835 \pm 6$  &  $165.000 \pm 83.000$  &  $6001 \pm 88$  &  ({\it 19--21}), This work  \\ 
\hline 
\end{tabular} 
} 

\noindent {\bf Table~S1.} {\bf Properties of all planetary systems with detections, non-detections, or upper-limit constraints of excess helium absorption.} The $\delta_{R_{p}}/H_{\rm eq}$ parameter expresses the equivalent height of the planets' helium upper atmospheres in units of their scale height at equilibrium temperature ($T_{\rm eq}$). The $R_{P}/R_{\rm RL}$ parameter expresses the Roche-lobe filling, computed as the ratio between the planetary radii and the radii of their Roche Lobes.

\newpage

\noindent \new{{\bf Data~S1.} {\bf Measured spectral and physical properties for individual HPF spectra of HAT-P-32~A+b.} This file includes, for each HPF spectrum, the file name, observing date and the corresponding orbital phase of the planet, barycentric and stellar RV, equivalent widths of the helium 10833~\AA\ feature and the calcium infrared triplet of HAT-P-32~A+b, central wavelength of the helium excess absorption feature of HAT-P-32~b, and the equivalent width of the planet's helium excess absorption.}

\end{document}